\documentclass[lettersize,journal]{IEEEtran}

\def\Ebb{{\mathbb E}}

\def\Gc{{\mathcal G}}

\def\Gbf{{\mathbf G}}

\def\Jc{{\mathcal J}}

\def\Lc{{\mathcal L}}

\def\Mc{{\mathcal M}}

\def\Nbb{{\mathbb N}}

\def\Rbb{{\mathbb R}}

\def\Uc{{\mathcal U}}

\def\0{{\bf 0}}

\def\underdotline#1{{\setbox0=\hbox{#1}\rlap{\raise-0.28em\hbox to\wd0{\rm\tiny\cleaders\hbox{.\kern-0.1ex}\hfill}}\box0}}
\def\underdashline#1{{\setbox0=\hbox{#1}\rlap{\raise-0.28em\hbox to\wd0{\rm\tiny\cleaders\hbox{-\kern-0.1ex}\hfill}}\box0}}
\def\understarline#1{{\setbox0=\hbox{#1}\rlap{\raise-0.28em\hbox to\wd0{\rm\tiny\cleaders\hbox{$\star$\kern-0.1ex}\hfill}}\box0}}

\newcommand{\bitem}{\begin{itemize}}
\newcommand{\eitem}{\end{itemize}}
\newcommand{\btabular}{\begin{tabular}}
\newcommand{\etabular}{\end{tabular}}
\newcommand{\bcenter}{\begin{center}}
\newcommand{\ecenter}{\end{center}}
\newcommand{\bea}{\begin{eqnarray}}
\newcommand{\eea}{\end{eqnarray}}
\newcommand{\bean}{\begin{eqnarray*}}
\newcommand{\eean}{\end{eqnarray*}}
\newcommand{\ba}{\left[ \begin{array}}
\newcommand{\ea}{\\ \end{array} \right]}
\newcommand{\bear}{\begin{array}}
\newcommand{\eear}{\\ \end{array}}

\newcommand{\non}{\nonumber}

\def\qed{{\ \vrule width 1.5mm height 1.5mm \smallskip}}

\font\myownfont=cmr17 scaled \magstep5
\def\psfancypar#1#2{\def\biginitial#1{{\myownfont#1}}%
  \def\makeinitial#1{\setbox8\hbox{\strut\vbox to 1.3ex
    {\hbox{\biginitial#1}\vskip -4pc plus 3.5pc minus 3.5pc}}}%
  \makeinitial#1%
  \ifdim\parindent>1.3\wd8\dimen8=\parindent
     \else\dimen8=1.3\wd8\fi
  \hangindent=\dimen8\hangafter=-2
  \noindent
  \strut\hskip-1\dimen8\box8{\sc#2}}%

\newcounter{subequation}
\def\beasub{\addtocounter{equation}{+1}
\setcounter{subequation}{\value{equation}}
\setcounter{equation}{0}
\renewcommand{\theequation}{\arabic{subequation}\alph{equation}}
\begin{eqnarray}}
\def\eeasub{\end{eqnarray}
\setcounter{equation}{\value{subequation}}
\renewcommand{\theequation}{\arabic{equation}}}

\newcommand{\overbar}[1]{\mkern 1.5mu\overline{\mkern-1.5mu#1\mkern-1.5mu}\mkern 1.5mu}

\newcommand{\bos}{\boldsymbol}

\newcommand{\bbm}{\begin{bmatrix}}
\newcommand{\ebm}{\end{bmatrix}}

\usepackage{amsmath,amsfonts} 
\usepackage{xcolor}
\usepackage{amssymb}
\usepackage{url}
\usepackage{graphicx}
\usepackage[export]{adjustbox}
\usepackage{array}
\usepackage{threeparttable}
\usepackage{multirow}
\usepackage{algorithmic}
\usepackage{mathtools}
\usepackage{algorithm}
\usepackage[shortlabels]{enumitem}
\usepackage[utf8]{inputenc}
\usepackage{textcomp}
\usepackage{stfloats}
\usepackage{url}
\usepackage{verbatim}
\usepackage{graphicx}
\usepackage{cite}
\usepackage[caption=false,font=footnotesize,labelfont=sf,textfont=sf]{subfig}
\usepackage{amsthm}
\usepackage{array,multirow}
\usepackage{multicol}
\newtheorem{theorem}{Theorem}
\newtheorem{lemma}[theorem]{Lemma}
\newtheorem{asst}{Assumption}
\newtheorem{rmk}{Remark}
\newtheorem{definition}{Definition}
\newcommand{\bpi}[1]{{\tilde{p}}_{#1,i}}

\newcommand{\xii}[1]{{#1}_{i,i}}
\newcommand{\ix}[2]{{#1}_{#2,i}}
\newcommand\myeq[1]{\mathrel{\stackrel{\makebox[0pt]{\mbox{\normalfont\tiny #1}}}{=}}}
\begin{document}
	
	\title{Asynchronous distributed collision avoidance with intention consensus for inland autonomous ships}

	\author{Hoang Anh Tran, Nikolai Lauvås, Tor Arne Johansen, Rudy R. Negenborn
        \thanks{This work has been submitted to the IEEE for possible publication. Copyright may be transferred without notice, after which this version may no longer be accessible.}
		 \thanks{The research leading to these results has received funding from the European Union's Horizon 2020 research and innovation programme under the Marie Skłodowska-Curie grant agreement No 955.768 (MSCA-ETN AUTOBarge), and the Researchlab Autonomous Shipping at Delft University of Technology. This publication reflects only the authors' view, exempting the European Union from any liability. Project website: http://etn-autobarge.eu/.}
		\thanks{Hoang Anh Tran, Nikolai Lauvås, and Tor Arne Johansen are with the Department of Engineering Cybernetics, Norwegian University of Science and Technology (NTNU), Norway.}
		\thanks{Rudy R. Negenborn is with the Department of Maritime and Transport Technology, Delft University of Technology, The Netherlands.} 
		\thanks{(e-mail: hoang.a.tran@ntnu.no, nikolai.lauvas@ntnu.no, tor.arne.johansen@ntnu.no, r.r.negenborn@tudelft.nl)}
	}
	\markboth{Journal of \LaTeX\ Class Files,~Vol.~14, No.~8, August~2021}%
	{Shell \MakeLowercase{\textit{et al.}}: A Sample Article Using IEEEtran.cls for IEEE Journals}
	
	\maketitle
	\begin{abstract}
		This paper focuses on the problem of collaborative collision avoidance for autonomous inland ships.
		Two solutions are provided to solve the problem in a distributed manner.
		We first present a distributed model predictive control (MPC) algorithm that allows ships to directly negotiate their intention to avoid collision in a synchronous communication framework.
		Moreover, we introduce a new approach to shape the ship's behavior to follow the waterway traffic regulations.
		The conditional convergence toward a stationary solution of this algorithm is guaranteed by the theory of the Alternating Direction Method of Multipliers (ADMM).
		To overcome the problem of asynchronous communication between ships, we adopt a new asynchronous nonlinear ADMM and present an asynchronous distributed MPC algorithm based on it.
		Several simulations and field experiments show that the proposed algorithms can prevent ship collisions even in complex scenarios.
	\end{abstract}
	
	\section{INTRODUCTION}
	The recent decade has witnessed an increase in research toward inland autonomous ships.
	One of the most critical components of an autonomous ship is the collision avoidance system (CAS), which ensures the ship's collision-free navigation.
	Several approaches have been proposed to increase the navigational safety of ships, such as MPC algorithms \cite{menges_nonlinear_2024,mahipala_model_2023}; Scenario-based model predictive control \cite{johansen16, trym_20}; Velocity obstacles \cite{zhang_22}.
	That being said, many of these studies are reactive methods that do not explicitly consider change of neighboring ships' intentions.

	There are two common approaches to dealing with the change in intentions of the neighboring ships.
	One approach is to predict the intentions of neighboring ships based on their current situation and/or their historical navigation data \cite{rothmund_22,rothmund_2023}.
	However, these algorithms cannot always guarantee high accuracy in the prediction.
	The other approach is to exchange intentions between ships through wireless communication.
	This approach removes the ambiguity of neighboring ships' intentions, but requires ships to be aided with sufficient communication equipment \cite{akdag22}.
	However, the rapid development of communication technology and protocols makes sharing of intentions between ships practical \cite{guiking_digital_nodate,projSTM}.
	
	There needs to be a framework for ships to collaborate efficiently when it comes to exchanging intentions between them.
	Different solutions have been proposed for a framework of collaborative collision avoidance (CCAS) of ships \cite{akdag_collaborative_2022,zheng17,du_22}.
	The CCAS framework can be categorized as centralized or distributed based on the architecture of the communication network.
	The difference between the two frameworks is the actor that executes the CCAS.
	In the centralized framework, the CCAS algorithm is performed by one specific coordinator, e.g., the traffic control center or one ship with powerful computation equipment, and the solution is sent to each ship \cite{chen_18,tam13,kurowski_19}.
	The ships are assumed to follow up precisely.
	This framework requires fewer onboard computational resources, but ships are vulnerable to communication faults, i.e., loss of communication with the coordinator.
	On the contrary, the distributed framework requires all ships to perform the CCAS algorithm individually and negotiate with each other to reach a consensus solution \cite{akdag_collaborative_2022, tran_parallel_2024,chen18}.
	In case of communication loss between ships, the CCAS algorithm onboard can act as a conventional CAS without intention sharing.
	Although the distributed framework requires more onboard computational resources, the robustness under unstable communication makes this framework more promising for future applications in CCAS.
	
	In CAS for autonomous vehicles, the MPC algorithm is widely used due to its ability to foresee and resolve the potential risks in complex traffic scenarios \cite{eriksen_20, yu_model_2021,ferranti_22}.
	A common approach to implement MPC algorithms in the distributed CCAS in autonomous ships is using the ADMM \cite{zheng17,ferranti18,chen_18}.
	The ADMM method was initially developed for the distributed optimization of a convex problem \cite{boyd_admm_2011}.
	However, the assumption of convexity may restrict the applications for the ADMM in CCAS.
	In \cite{themelis_20}, Themelis et al. presented a universal equivalent between the nonconvex Douglas-Rachford splitting method and ADMM for a class of nonconvex optimization problems.
	This finding not only allows for the direct application of nonlinear ADMM (NADMM) in CCAS \cite{ferranti_22,tran_distributed_2024} but also opens an opportunity to apply the theory of Douglas-Rachford splitting to NADMM algorithms.
	
	When using ADMM to solve the distributed CCAS problem, one important factor is whether the communication is synchronous or asynchronous.
	In a synchronous communication network, controllers must wait for each other to acknowledge the new update before continuing.
	On the other hand, a controller in an asynchronous communication network can send updates continuously without waiting for others.
	Although the distributed ADMM attains linear and global convergence in synchronous networks \cite{shi2014linear,makhdoumi2017convergence}, the synchronous communication in CCAS for autonomous ships is either hard to implement or can delay the process.
	Several approaches have been proposed to tackle the problem of asynchronous convex ADMM \cite{bastianello_21,pmlr-v32-zhange14,chang2016asynchronous}.
	However, there are limited results for the nonconvex ADMM problems.
	In \cite{hong2017distributed} and \cite{wang_fedadmm_2022}, two asynchronous ADMM approaches for nonconvex problems for a specific type of equality constraint are presented.
	However, our paper's formulated CCAS problem uses a more general equality constraint and extends the aforementioned algorithms.
	%
	
	This paper presents two approaches to solving the problem of distributed CCAS for autonomous ships in inland waterways:
	\begin{enumerate}
		\item \textit{The synchronous CCAS (Sync-CCAS)}, a distributed MPC for the CCAS problem considering synchronous communication between ships. In addition, this algorithm explicitly considers inland waterway traffic regulations.
		\item \textit{The asynchronous CCAS (Async-CCAS)}, which has the same properties as the Sync-CCAS algorithm, but concerns the case of asynchronous communication, with signal delays between ships.
	\end{enumerate}
	
	%
	%
	%
	We consider the set $\Mc$ of $M$ ships to solve the CAS problem collaboratively in a distributed manner.
	The consensus of intentions between ships is reached through negotiation using digital ship-to-ship communication.
	
\begin{figure}[!t]
		\centering
		\includegraphics[width=0.8\linewidth]{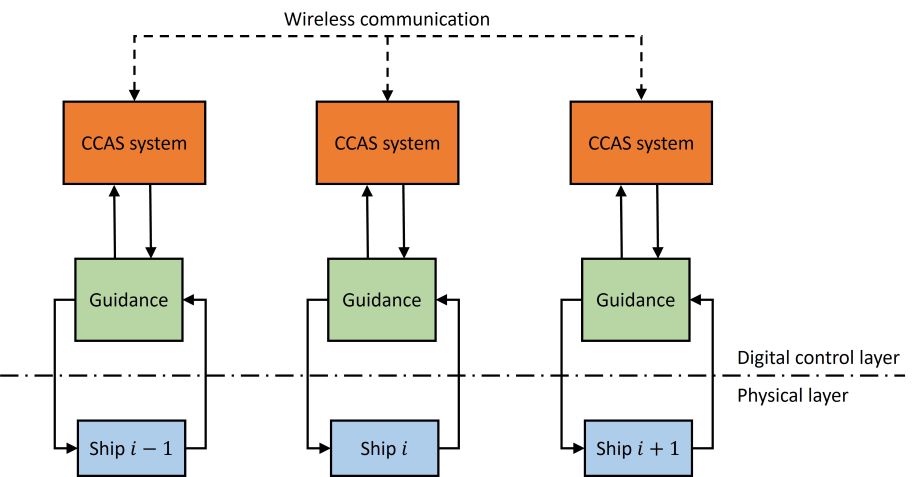}
		\caption{Collaborative collision avoidance scheme between ships}.
		\label{CAS-protocol}
	\end{figure}

	Fig. \ref{CAS-protocol} shows the control scheme for autonomous ship $i$ with the CCAS in the loop.
	The guidance system keeps the ship on track with the predefined waypoints by commanding the control system to maintain the desired speed and course angle.
	The CCAS continuously checks for potential collision risks on the sailing path.
	If a risk of collision is detected, the CCAS will perform the necessary course change and/or speed change modification to avoid the collision and achieve compliance with traffic rules when possible.
	Besides, the CCAS also exchanges the predicted trajectories with neighboring ships.
	The CCAS includes the following three main components:
	\begin{enumerate}
		\item \textit{Kinematic model of ships:} To predict the future position of the own and the neighboring ships.
		\item \textit{Collision risk evaluation:} To evaluate the situation and detect potential collision risks and violations of traffic rules based on predicted future trajectories of all ships.
		\item \textit{The CCAS algorithm:} To decide the necessary evasive maneuver based on the potential risks.
	\end{enumerate}
	We adopt the NADMM in \cite{themelis_20} to develop the synchronous CCAS algorithm, where its preliminary version was presented in \cite{tran_parallel_2024}.
	Compared to \cite{tran_parallel_2024}, we introduce a new approach to optimize ship behavior following waterway traffic regulations.
	The new approach is better at handling cases where neither ship has a duty to give way and needs to negotiate to avoid collision.
	Moreover, this paper provides a theoretical analysis of the convergence of the synchronous algorithm.
	To deal with asynchronous communication, we present an asynchronous NADMM (Async-NADMM) algorithm for nonconvex optimization problems with a relaxed equality constraint.
	Due to the equivalence between nonconvex Douglas-Rachford splitting method and NADMM, we can cast the result of the asynchronous Douglas-Rachford splitting method in \cite{tran_dinh_feddr_2021} to the NADMM problem.
	Then, the Async-CCAS is formed based on the Async-NADMM.
	The performance of the proposed algorithms is verified in both simulation and field experiments with various representative traffic scenarios.
	
	The remainder of this paper is structured as follows.
	Section \ref{sec:pre} presents preliminary information needed to formulate the CCAS in the later section, including notations, definitions, and the components of the CCAS.
	The Sync-CCAS and its convergence analysis are presented in Section \ref{sec:sync-COLAV}.
	Section \ref{sec:async-COLAV} introduces the Async-ADMM and, with it, the Async-CCAS.
	Finally, the experimental results are presented in Section \ref{sec:sim}, and Section \ref{sec:conclusion} concludes the paper.
	
	\section{Preliminaries}\label{sec:pre}
	This section presents notations and definitions that will be used.
	We also briefly present the first two components of the CCAS that provide the background for the proposed CCAS algorithms.
	Additionally, this section presents the NADMM, which is used as a framework for the Sync-CCAS.
	\subsection{Notations and definitions}
	We denote $\textbf{range}(M)$ as the range (column space) of the matrix $M$.
	%
	The domain of an extended-real-valued function $f: \Rbb^n \rightarrow \overbar{\Rbb}$ is $\textbf{dom} (f) := \left\{x \in \Rbb^n~|~f(x) < \infty \right\}.$
	\begin{definition}[lower semicontinuous function]
		A function $f: X \rightarrow \overbar{\Rbb}$ is called lower semicontinuous (lsc) at point $x_0 \in X$ if $\lim \inf_{x \rightarrow x_0} f(x) = f(x_0)$. Furthermore, $f$ is called lsc if $f(x)$ is lsc at every point $x_0 \in \textbf{dom}(f)$.
	\end{definition}
	\begin{definition}[Lipchitz continuous gradient]
		A differentiable function $h$ is said to have Lipschitz continuous gradient with constant $L_h>0$ (or $L_{h}-\text{smooth}$) on $\textbf{dom}(h)$ if 
		\begin{align*}
			||\nabla h(x_1)- \nabla h(x_2)|| \leq L_h ||x_1 -x_2||, ~~ \forall x_1,x_2 \in \textbf{dom}(h).
		\end{align*}
	\end{definition}
	\begin{definition}[Stationary point]
		Given $f: \Rbb^n \rightarrow \overbar{\Rbb}$. Then $x^* \in \textbf{dom}(f)$ is called an stationary point of $f(x)$ if $ \nabla f(x^*) =0$.
	\end{definition}
	%
	\begin{definition}[image function]
		Given $f: \Rbb^n \rightarrow \overbar{\Rbb}$ and $M \in \Rbb^{m \times n}$. Then the image function $(Mf): \Rbb^m \rightarrow [-\infty,+\infty]$ is defined as $(Mf)(\epsilon) := \inf_{x \in \Rbb^n}\left\{f(x)~|~Mx=\epsilon \right\}$.
	\end{definition}
	\begin{definition}[Proximal mapping]
		Given $f: \Rbb^n \rightarrow \overbar{\Rbb}$ and $\gamma >0$. Then the proximal mapping $\textbf{prox}_{\gamma f}: \Rbb^n \rightarrow \textbf{dom}(f)$ is defined as:
		\begin{align}\label{prox}
			\textbf{prox}_{\gamma f}(x) = \arg \min_y\left\{f(y)+\frac{1}{2\gamma}||y-x||^2\right\}.
		\end{align}
	\end{definition}
	If $f$ is $L_f$-smooth and $\gamma<L_f$, then $\textbf{prox}_{\gamma f}(x)$ is well-defined and single valued.
	Furthermore, as pointed out in \cite{themelis_20}, we also have the following property:
	\begin{align}\label{prox-con}
		\frac{y-x}{\gamma} \in \nabla f(x), ~~~~\forall x = \textbf{prox}_{\gamma f}(y).
	\end{align}
	
\begin{figure}[!t]
		\centering
		\includegraphics[width=0.8\linewidth]{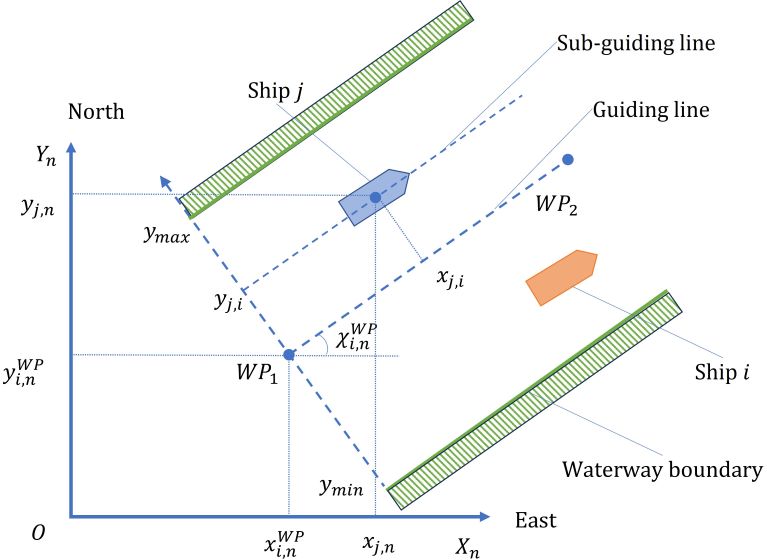}
		\caption{Path coordinate frame $\{\varpi_i\}$, with origin at WP1, and inertial coordinate frame $\{n\}$.}
		\label{coordinate}
	\end{figure}

	In this paper, the path coordinate frame $\{\varpi_i\}$ is used by ship $i\in \Mc$ to represent the state of itself and neighboring ships.
	We denote $p_{j,i} = [x_{j,i}, y_{j,i}, \chi_{j,i}]^\top$ as the state of ship $j$ with respect to coordinate frame $\{\varpi_i\}$, consisting of coordinates along and across guideline between waypoints, and the angle relative to guideline.
	%
	The relation between $p_{j,i}$ and the state of ship $j$ in the inertial coordinate frame $\{n\}$, i.e., $\eta_j = [x_{j,n},y_{j,n}, \chi_{j,n}]^\top$,  is as follows:
	\begin{align*}
		p_{j,i} &=\bbm \ix{x}{j} \\ \ix{y}{j} \\ \ix{\chi}{j} \ebm = R_i \bbm x_{j,n} - x_{i,n}^\mathrm{WP} \\ y_{j,n}- y_{i,n}^\mathrm{WP}  \\ \chi_{j,n} - \chi_{i,n}^\mathrm{WP}  \ebm = R_i (\eta_j-{\eta}_{i}^\mathrm{WP}),\\
		R_i &=\bbm \cos\left(\chi_{i,n}^\mathrm{WP}\right) & \sin\left(\chi_{i,n}^\mathrm{WP}\right) & 0 \\
		-\sin\left(\chi_{i,n}^\mathrm{WP}\right) & \cos\left(\chi_{i,n}^\mathrm{WP}\right)  & 0 \\
		0& 0 & 1\ebm,
	\end{align*} 
	where ${\eta}_{i}^\mathrm{WP}=[x_{i,n}^\mathrm{WP},y_{i,n}^\mathrm{WP}, \chi_{i,n}^\mathrm{WP}]^\top$ is the parameter of the previous active waypoint in the inertial frame (see Fig. \ref{coordinate}).
	\subsection{Kinematic model of ships}
	
	\subsubsection{Kinematic model of own-ship}
	The kinematic model of ship $i$ with respect to coordinate frame $\{\varpi_i\}$ is described according to \cite{tran_collision_2023} as follows:
	\begin{align}
		\begin{split}
			\xii{x}(k+1) &= \xii{x}(k) + u_{i,i}^s(k)U_i^d\cos(\xii{\chi}(k))\Delta T,\\
			\xii{y}(k+1) &= \xii{y}(k) + u_{i,i}^s(k)U_i^d\sin(\xii{\chi}(k))\Delta T,\\
			\xii{\chi}(k+1) &= \xii{\chi}(k)  
			+ \frac{\Delta T}{T_{1}} \cdot \\ 
			& \Big[ \chi^{\max}_i\tanh(K_e (u_{i,i}^y(k) - \xii{y}(k))) - \xii{\chi}(k)\Big],
		\end{split}
		\label{kinematic_model}
	\end{align}
	where $U_i^d$ is the nominal surge speed; $\chi^{\max}_i$ is the maximum course angle change that ship $i$ can achieve within a sampling period $\Delta T$; $K_e$ and $T_1$ are the tuning parameter of the guidance law, in this case the line of sight guidance law, and the positive constant influenced by the properties of the autopilot system and ship hydrodynamics.
	The control inputs of model \eqref{kinematic_model} are the cross-track offset $u_{i,i}^y$, and speed modification $u_{i,i}^s$.
	%
	%
	%
	Hence, $u_{i,i}(k) = [u_{i,i}^y(k), u_{i,i}^s(k))]^\top$ denotes the vector of the control signal of ship $i$.
	\subsubsection{Kinematic model of neighboring ships}
	The kinematic model of ship $j$ with respect to the coordinate frame $\{\varpi_i\}$ can be written as follows:
	\begin{align}
		\ix{x}{j}(k+1) &= \ix{x}{j}(k) + u_{j,i}^s(k) U_j(k)\cos(\ix{\chi}{j}(k))\Delta T,\nonumber\\
		\ix{y}{j}(k+1) &= \ix{y}{j}(k) + u_{j,i}^s(k)U_j(k)\sin(\ix{\chi}{j}(k))\Delta T,\label{kinematic_ns}\\
		\ix{\chi}{j}(k+1) &= \ix{\chi}{j}(k) + \frac{\Delta T}{T_j}  \left(\ix{\chi}{j}^d(k)+\ix{\chi}{j}^\mathrm{nom}-\ix{\chi}{j}(k) \right) \nonumber,
	\end{align}
	where $U_j$ is the nominal surge speed of ship $j$, and $T_j$ is the time constant of the course dynamics.
	Additionally, $\ix{\chi}{j}^\mathrm{nom} = \chi_{j,n}^\mathrm{WP} - \chi_{i,n}^\mathrm{WP}$ is the angle of the active reference course of ship $j$, i.e., $\chi_{j,n}^\mathrm{WP}$, with respect to coordinate frame $\{\varpi_i\}$. 
	The vector of input signals is $u_{j,i}(k) = [u_{j,i}^s(k),\ix{\chi}{j}^d(k)]^\top$, where $u_{j,i}^s(k)$ and $\ix{\chi}{j}^d(k)$ are the speed and course adjustments.
	In this paper, we assume that ship $i$ gets the information of the model \eqref{kinematic_ns} of ship $j$ through communication.
	
	For sake of notational simplicity, let us rewrite models \eqref{kinematic_model} and \eqref{kinematic_ns}, respectively, as follows:
	\begin{align*}
		p_{i,i}(k+1) = f_{i,i}(p_{i,i}(k),u_{i,i}(k)),\\
		p_{j,i}(k+1) = f_{j,i}(p_{j,i}(k),u_{j,i}(k)).
	\end{align*}
	
	\subsection{Collision risk evaluation}
	%
	The prediction of the collision risk of ship $i$ concerning ship $j$ at $k$ time steps from the present time $t_0$ is $R_{ij}(t_0+k)$, described as follows:
	\begin{align*}
		R_{ij}(t_0+k) &= \frac{K_\mathrm{ca}}{\sqrt{1+K_\mathrm{d} k}} D_x(t_0+k)D_y(t_0+k), \\
		D_x(t_0+k) &= exp\left[-\frac{\left(\ix{x}{i}(t_0+k) - \ix{x}{j}(t_0+k)\right)^2}{\alpha_{xj}}\right],\\
		D_y(t_0+k) &= exp\left[-\frac{\left(\ix{y}{i}(t_0+k) - \ix{y}{j}(t_0+k)\right)^2}{\alpha_{yj}}\right],
	\end{align*}
	where $K_\mathrm{ca}$ is a predefined constant determined by safety criteria, which varies according to the traffic situation; $\frac{1}{\sqrt{1+K_\mathrm{d} k}}$, with $K_d \geq 0$ being a discount factor that reduces the collision risks associated with larger $k$.
	Additionally, $\alpha_{xj}$ and $\alpha_{yj}$ are parameters characterizing the size and shape of ship $j$.
	\subsection{Nonlinear ADMM}
	Given the optimization problem as follows:
	\begin{subequations}\label{ADMM-prob}
		\begin{align}
			\min_{(w,v) \in \Rbb^m \times \Rbb^n}  & \quad  f(w) +g(v), \\
			s.t. 				 & \quad A w -Bv -c = 0,
		\end{align}
	\end{subequations}
	where  $f: \Rbb^m \rightarrow \overbar{\Rbb}$, $g: \Rbb^n \rightarrow \overbar{\Rbb}$, $A \in \Rbb^{q \times m}$, $B \in \Rbb^{q \times n}$, and $c \in \Rbb^q$, the augmented Lagrangian of problem \eqref{ADMM-prob} with the Lagrange multiplier $z\in \Rbb^q$ is:
	\begin{align*}
		\Lc_{\beta}(w,v,z) &= f(w) + g(v) + \langle z, Aw - Bv -c\rangle \\
		&+\frac{\beta}{2}||Aw - Bv -c||^2
	\end{align*}
	Then, the NADMM algorithm in \cite{themelis_20} solves problem \eqref{ADMM-prob} iteratively as:
	\begin{align}\label{NADMM}
		\begin{cases}
			z^{+1/2} &= z - \beta(1-\lambda)(Aw - Bv -c),\\
			w^+      &\in \arg\min \Lc_{\beta}(\cdot,v,z^{+1/2}),  \\
			z^{+}    &= z^{+1/2} + \beta(Aw - Bv -c),\\
			v^+      &\in \arg\min \Lc_{\beta}(w,\cdot,z^{+}),
		\end{cases}
	\end{align}
	where $\beta>0$, and $\lambda \in (0, 2)$ are tuning parameters.
	%
	\section{Synchronous distributed collaborative collision avoidance algorithm} \label{sec:sync-COLAV}
	This section presents the Sync-CCAS algorithm.
	The Sync-CCAS is developed based on the assumption of synchronous communication.
	That is, after every negotiation round, each ship waits for the decisions of all neighboring ships before starting the next round.
	%
	\subsection{Problem formulation}
	The problem of CCAS for autonomous ships is formulated as a distributed MPC problem.
	The cost function combines the potential risk and the cost of taking action.
	By optimizing the cost function, we aim to find a safe trajectory that complies with traffic rules, and with minimum effort.

	For simplicity, we denote $\tilde{u}_{i,i} =[u_{i,i}^\top(t_0), u_{i,i}^\top(t_0+1),...,u_{i,i}^\top(t_0+N-1)]^\top$, $\bpi{i} = [p_{i,i}^\top(t_0), p_{i,i}^\top(t_0+1),...,p_{i,i}^\top(t_0+N)]^\top$ as, respectively, control input and state vectors of ship $i$ over a control horizon of $N$ time steps.
	Moreover, each ship $i\in \Mc$ stores a local copy of the position of all neighboring ships with respect to the coordinate frame $\{\varpi_i\}$, denoted as $\tilde{p}_i = [\bpi{1}^\top,\bpi{2}^\top,...,\bpi{M}^\top]^\top$.
	Similarly, $\tilde{u}_i = [\tilde{u}_{1,i}^\top,\tilde{u}_{2,i}^\top,...\tilde{u}_{i,i}^\top,..\tilde{u}_{M,i}^\top]^\top$ is ship $i$'s prediction of the control input of all ships in $\Mc$.
	%
	%
	Besides the local variable $\tilde{p}_i$, we define $\xi$ as the global variable that stores the consensus trajectory of all ships over $N$ time steps.
	The relation between the local and the global variable is
	%
	\begin{align}
		\tilde{p}_i = \tilde{R}_i (\xi - \bar{\eta}_{i}^\mathrm{WP}), \quad \forall i \in \Mc,
	\end{align}
	where $\bar{\eta}_{i}^\mathrm{WP} = \bos{1}_{M(N+1)} \otimes {\eta}_{i}^\mathrm{WP}$ with $\otimes$ denoting the Kronecker product, and $\tilde{R}_i$ is defined as follows:
	\begin{align*}
		\tilde{R}_i = \bbm R_i & &0 \\
		&\ddots&\\
		0 & & R_i\ebm \in \Rbb^{3(N+1)M\times 3(N+1)M}.
	\end{align*}

	Then, we formulate the cost function for the MPC problem of ship $i$ as follows:
	\begin{align}
		\Jc_i(\tilde{p}_i,\tilde{u}_{i}) =\Jc_i^{ca}(\tilde{p}_i) + \Jc_i^{e}({\tilde{u}}_{i}) + \Jc_i^{b}({\tilde{u}}_{i}).
		\label{cost}
	\end{align}
	Here $\Jc_i^{ca}(\tilde{p}_i) = \sum_{k=1}^{N+1} \sum_{j \in \Mc \backslash \{i\}} R_{ij}(t_0+k)$ is the cumulative sum of risk functions with all neighboring ships over the horizon.
	$\Jc_i^{e}({\tilde{u}}_{i})$ is the cost of control actions, defined as:
	%
	%
	\begin{align} \label{cost_u}
		\Jc_i^{e}({\tilde{u}}_{i}) &= \sum_{k=1}^{N+1} \left[K_y\left(u^y_{i,i}(t_0+k) - u^y_{i,i}(t_0+k-1)\right)^2 \right. \non\\
		&\left.+ K_s\left(1-u^s_{i,i}(t_0+k)\right)^2 \right] \\
		& + \sum_{j\in \Mc \backslash \{ i\} }\sum_{k=1}^{N+1} \alpha_{j,i}\big[ (\chi_{j,i}^d(k))^2 + (1-u_{j,i}^s(k))^2 \big], \non
	\end{align}
	where $K_y$, $K_s$ are positive tuning parameters.
	The first term in \eqref{cost_u} is the cost of ship $ i$'s action and indicates that ship $i$ should only change its course if it can significantly lower the risk of a collision.
	The second term in \eqref{cost_u} is the cost of proposing course changes to neighboring ships.
	Depending on the priority constant $\alpha_{j,i}>0$, ship $i$ can decide to change course or propose ship $j$ to change course.
	Details on choosing appropriate $\alpha_{j,i}$ are discussed in Section \ref{tapd}.
	Finally, $\Jc_i^{b}({\tilde{u}}_{i})$ is a term that makes the behavior of ship $i$ adequately represent waterway traffic regulations, e.g. preferring to steer towards starboard in a head-on situation.
	More details on $\Jc_i^{b}({\tilde{u}}_{i})$ can be found in \cite{tran_distributed_2024}.
	%
	
	
	We formulate the distributed MPC collision avoidance problem of ship $i \in \Mc$, with cost function \eqref{cost} as follows:
	\begin{subequations}\label{prob1}
		\begin{align}
			\min_{\tilde{p}_i,{\tilde{u}}_i}  & \quad \Jc_i(\tilde{p}_i,\tilde{u}_{i}) \\
			\textrm{s.t.:}   &  \quad  p_{i,i}(t_0+k+1) = f_{i,i}(p_{i,i}(t_0+k),u_{i,i}(t_0+k)),  \label{con_p1}\\
			&  \quad p_{j,i}(t_0+k+1) = f_{j,i}(p_{j,i}(t_0+k),u_{j,i}(t_0+k)), \label{con_pji}\\
			&  \quad  p_{j,i}(t_0) = p_{j,i}^\mathrm{init},  \label{con_p2}\\
			&  \quad  {u}_{i,i}(t_0+k) \in \Uc_i, \label{con_u} \\
			& 	\quad u_{j,i}(t_0+k) \in \Uc_{j,i} \label{con_uj}\\
			&  \quad \tilde{p}_i =\tilde{R}_i (\xi - \bar{\eta}_{i}^\mathrm{WP} ),  \label{con1a}
		\end{align}
	\end{subequations}
	$\forall j \in \Mc$, where $\Uc_i$ and $\Uc_{j,i}$ are the admissible sets of $u_{i,i}$ and $u_{j,i}$ respectively.
	To ensure that ship $i$ cannot propose another ship to steer port side, we bound $\ix{\chi}{j}^d(k)$ below by $0$.
	
	To fit problem \eqref{prob1} into the NADMM framework, we define $\Gbf_i:=\left\{[\tilde{u}_i^\top,\tilde{p}_i^\top]^\top|\text{\eqref{con_p1}, \eqref{con_pji}, \eqref{con_p2}, \eqref{con_u}, \eqref{con_uj} are satisfied}\right\}$ as the feasible region for states and inputs of ship $i$.
	%
	%
	%
	%
	Then, for ship $i \in \Mc$, the NADMM update at iteration with index $s$ is defined as follows:
	\begin{subequations} \label{ADMM-update}
		\begin{align}
			\xi^{s+1} &= \frac{1}{M} \sum_{j=1}^{M} \hat{\xi}_j^{s} \label{update_global}\\
			z^{s+1/2}_{i} &= z^{s}_{i} - \beta(1-\lambda)\left(\tilde{R}^{-1}_i\tilde{p}_i^{s} + \bar{\eta}_{i}^\mathrm{WP}- \xi^{s+1}\right), \label{update_z1}\\
			\bbm \tilde{u}_i^{s+1} \\ \tilde{p}_i^{s+1}\ebm &= \text{argmin}_{ [\tilde{u}_i^\top,\tilde{p}_i^\top]^\top \in \Gbf_i} \left\{\Lc_i(\tilde{p}_i,\tilde{u}_i)\right\}, \label{update_local}\\
			z^{s+1}_{i} &= z^{s+1/2}_{i} + \beta\left(\tilde{R}^{-1}_i\tilde{p}_i^{s+1} + \bar{\eta}_{i}^\mathrm{WP}- \xi^{s+1}\right) , \label{update_z2}\\
			%
			%
			\hat{\xi}_j^{s+1} &=\tilde{R}^{-1}_i\tilde{p}_i^{s+1} + \bar{\eta}_{i}^\mathrm{WP} + \frac{1}{\beta}z^{s+1}_{i}, \label{update_global_local}
		\end{align}
	\end{subequations}
	where 
	\begin{align*}
		\Lc_i(\tilde{p}_i,\tilde{u}_{i}) = &\Jc_i(\tilde{p}_i,\tilde{u}_{i}) \\&+ \left<z_i^{s+1/2},\left(\tilde{R}^{-1}_i\tilde{p}_i + \bar{\eta}_{i}^\mathrm{WP}- \xi^{s+1}\right)\right>\\
		&+ \frac{\beta}{2}\Bigl|\Bigl|\left(\tilde{R}^{-1}_i\tilde{p}_i + \bar{\eta}_{i}^\mathrm{WP}- \xi^{s+1}\right)\Bigr|\Bigr|^2,
	\end{align*}
	with $z_i$ being the vector of Lagrange multipliers.
	The detailed steps of the asynchronous collaborative collision avoidance algorithm are presented in Algorithm \ref{sync-CCAS}.
	It is worth noting that if there is a change in the number of participant ships in Algorithm \ref{sync-CCAS}, e.g., a new ship enters the traffic situation, then the problem \eqref{prob1} needs to be reformulated.
	\begin{algorithm}[!t]
		\caption{Sync-CCAS} \label{sync-CCAS}
		\begin{algorithmic}[1]
			\FOR{ $s =1,..., s_{\max}$}
			\FORALL{$i \in \Mc$ in parallel}	
			\STATE Receive $\hat{\xi}_j^{s}$ from neighboring ships
			\STATE Update the global variable $\xi^{s+1}$ using \eqref{update_global}
			\STATE Update local variables $z^{s+1}_{i}$, $\tilde{u}_i^{s+1}$, and  $\tilde{p}_{i}^{s+1}$ using \eqref{update_z1}--\eqref{update_global_local} \label{sync:updateADMM}
			\STATE $\hat{\xi}_j^{s+1}=\tilde{R}^{-1}_i\tilde{p}_i^{s+1} + \bar{\eta}_{i}^\mathrm{WP}$ \label{sync:transform}
			\STATE Transmit data $\hat{\xi}_i^{s+1} $ to all ship $j \in \Mc \backslash \{ i \}$. \label{sync:transmit}	
			\ENDFOR
			\STATE Wait for the update of $\hat{\xi}_j^{s+1} $ from all neighboring ships.
			\STATE $s := s+1 $.
			\ENDFOR
		\end{algorithmic}
	\end{algorithm}
	
	\subsection{Choosing $\alpha_{j,i}$ based on traffic regulations}
	\label{tapd}
\begin{figure}[!t]
		\centering
		\subfloat[]
		{\centering
			\includegraphics[width=0.3\linewidth]{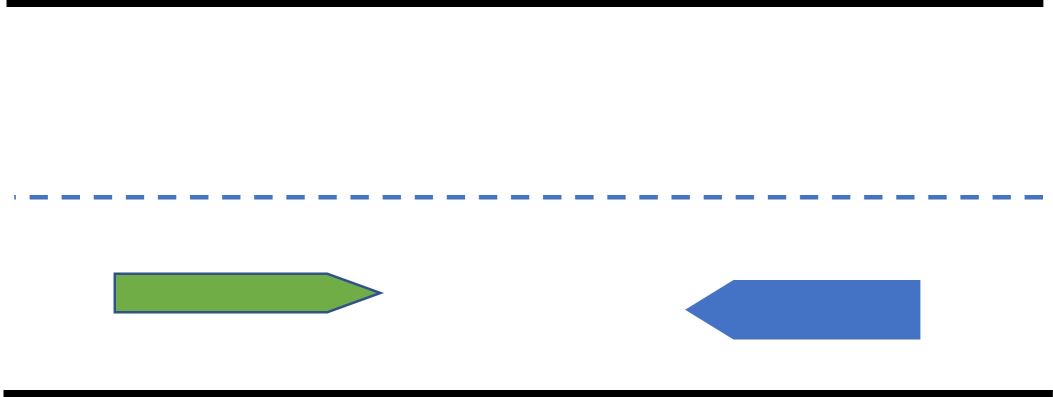}
			\label{headon1}}
		\hfil
		\subfloat[]
		{\centering
			\includegraphics[width=0.33\linewidth]{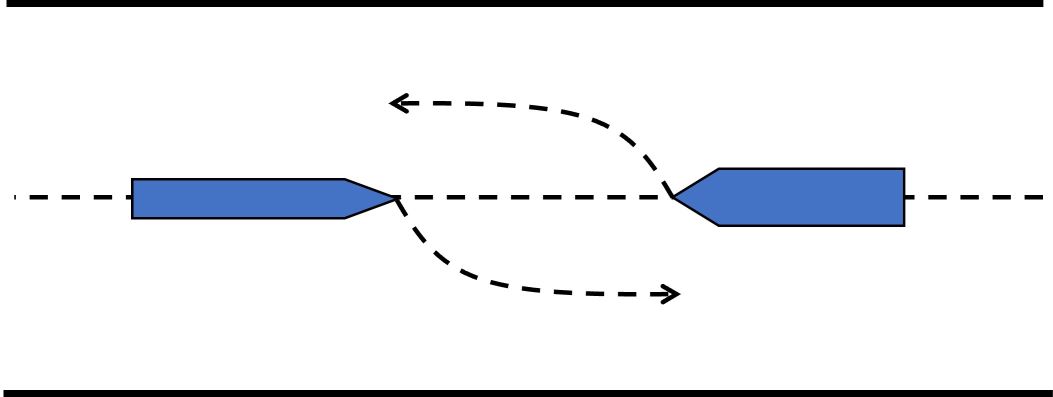}
			\label{headon2}}
		\hfil
		\centering
		\subfloat[]
		{\centering
			\includegraphics[width=0.33\linewidth]{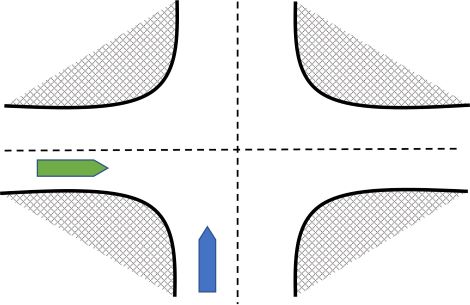}
			\label{cross1}}
		\hfil
		\subfloat[]
		{\centering
			\includegraphics[width=0.33\linewidth]{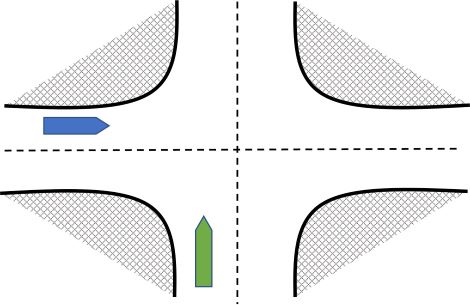}
			\label{cross2}}
		\hfil
		\caption{Traffic situations according to the Netherlands' inland shipping police regulations: (a), (b) head-on situations; (c), (d) crossing situations. The blue ship is the give-way ship, and the green is a stand-on ship.}
	\end{figure}

	From \eqref{cost_u}, if we choose $\alpha_{j,i}$ large enough, then the cost for ship $i$ to propose a course change for ship $j$ will be prohibitive.
	Therefore, in this case, ship $i$ will change its course instead of proposing a course change to $j$ and behave as a give-way ship.
	Conversely, if we choose $\alpha_{j,i}$ small enough, ship $i$ will behave as a stand-on ship by keeping its course and require ship $j$ to change course.
	However, if $\alpha_{j,i}$ is neither too small nor too large, e.g., $\alpha_{j,i}=1$, then behavior between ship $i$ and $j$ is decided through the negotiation process, i.e., the update iteration of NADMM.
	The stand-on or give-way priority between ship $i$ and $j$ is determined based on traffic regulations.
	This paper uses traffic regulations adopted in Chapter 6 of the Netherlands' inland shipping police regulations \cite{binnen}.
	The following rules are considered:
	\bitem
	\item Head-on situation: If two vessels are approaching each other on opposite courses in such a way that there is a risk of collision, the vessel not following the starboard side of the fairway shall give way to the vessel following the starboard side of the fairway (see Fig. \ref{headon1}). If neither vessel follows the starboard side of the fairway, each shall give way to vessels on the starboard side so that they pass each other port to port (see Fig. \ref{headon2}).
	\item Crossing situation: If the courses of two ships cross each other in such a way that there is a risk of collision, the vessel not following the starboard side of the fairway shall give way to the vessel following the starboard side of the fairway (see Fig. \ref{cross1}). In case none of the ships follows the starboard side of the fairway, the ship approaching from the port side gives way to the vessel approaching from starboard (see Fig. \ref{cross2}).
	\item Overtaking situation: A vessel overtaking another vessel should keep out of the way of the overtaken vessel.
	\eitem
	Consequently, we choose $\alpha_{j,i}$ as follows:
	\bitem
	\item If ship $i$ shall stand on to ship $j$, then $\alpha_{j,i}=K_\mathrm{SO}$, with $K_\mathrm{SO}\in (0,1)$ small enough.
	\item If neither of ship $i$ nor $j$ has stand on priority over each other then $\alpha_{j,i}=1$.
	\item If ship $i$ shall give way to ship $j$, then $w_{ji}=K_\mathrm{GW}$, with $K_\mathrm{GW}>1$ large enough.
	\eitem
	\begin{rmk}
		The different values of $\alpha_{j,i}$ could change the outcome of the solution provided by Sync-CCAS.
		Therefore, $\alpha_{j,i}$ is determined before the NADMM start and kept unchanged during the negotiation process, even if there is a change in the priority between ships.
	\end{rmk}

	\subsection{Convergence analysis of Algorithm \ref{sync-CCAS}}
	First, let us rewrite the collaborative collision avoidance problem \eqref{prob1} in the following standard form:
	\begin{subequations}\label{prob2}
		\begin{align}
			\min_{w,v}  & \quad  F(w,v) =\frac{1}{M}\sum_{i \in \Mc}f_i(w_i) +g(v)\\
			\textrm{s.t.:}   &  \quad  A_iw_i - v - c_i = 0, \ \  in \in \Mc
		\end{align}
	\end{subequations}
	where
	\begin{subequations}
		\begin{align}
			w_i &= [\tilde{u}_i^\top, \tilde{p}_{i}^\top]^\top, \\
			v &= \xi,\\
			f_i(w_i) &= \Jc_i(\tilde{p}_{i} ,\tilde{u}_{i}) + \Gc_i(\tilde{p}_{i} ,\tilde{u}_i), \label{indicator}\\
			g(\xi)&=0\\
			A_i &=\bbm 0_{3M(N+1) \times 2MN} & \tilde{R}^{-1}_i\ebm,\\
			c_i &=-\bar{\eta}_{i}^\mathrm{WP},
		\end{align}
	\end{subequations}
	and $\Gc_i(\tilde{p}_{i} ,\tilde{u}_i)$ is an indicator function of the feasible region $G_i$, i.e., $\Gc_i(\tilde{p}_{i} ,\tilde{u}_i)=0$ if $({\tilde{p}_{i} ,\tilde{u}_i}) \in G_i$ and $\Gc_i(\tilde{p}_{i} ,\tilde{u}_i)=\infty$ otherwise.
	
	It is worth noting that if we combine the update \eqref{update_global_local} and \eqref{update_global} as follows:
	\begin{align*}
		\xi^{s+1} = \frac{1}{M} \sum_{j=1}^{M}\left\{\tilde{R}^{-1}_i\tilde{p}_i^{s+1} + \bar{\eta}_{i}^\mathrm{WP} + \frac{1}{\beta}z^{s+1}_{i}\right\},
	\end{align*}
	and let
	\begin{align*}
		w &= [w_1^\top,w_2^\top,...,w_M^\top]^\top,\\
		z &= [z_1^\top,z_2^\top,...,z_M^\top]^\top,\\
		f(w) &= \frac{1}{M}\sum_{i \in \Mc}f_i(w_i),\\
		A &= \bbm A_1 & 0   & \cdots & 0 \\
		0   & A_2 & \cdots & 0 \\
		\vdots &\vdots & \ddots &\vdots\\
		0   & 0   & \cdots & A_M \ebm,\\
		B &=  \bbm I_{3MN} & I_{3MN} &\cdots&I_{3MN} \ebm^\top,
	\end{align*}
	then the NADMM update \eqref{ADMM-update} becomes:
	\begin{subequations} \label{ADMM-update-2}
		\begin{align}
			z^{s+1/2} &= z^{s} - \beta(1-\lambda)\left(Aw^{s} - c- Bv^{s}\right)\\
			w^{s+1}  &= \text{argmin}_{w} \Bigg\{ f(w)+ \langle z^{s+1/2}, Aw - Bv^{s} -c\rangle \nonumber\\
			& ~~~~~~~~~~~~~~~+\frac{\beta}{2}||Aw - Bv^{s} -c||^2 \Bigg\},\label{update_local-2}\\
			z^{s+1}_{i} &= z^{s+1/2}_{i} + \beta\left(Aw^{s+1} - c- Bv^{s}\right),\\
			v^{s+1} &= \frac{1}{M} \sum_{j=1}^{M}\left\{A_iw_i^{s+1} -c_i + \frac{1}{\beta}z^{s+1}_{i}\right\} \nonumber\\
			&=\text{argmin}_{v} \Big\{\langle z^{s+1/2}, Aw^{s+1} - Bv -c\rangle \nonumber\\
			& ~~~~~~~~~~~~~~~+\frac{\beta}{2}||Aw^{s+1} - Bv -c||^2 \Big\}.
		\end{align}
	\end{subequations}
	Here, we get update \eqref{update_local-2} by stacking the update \eqref{update_local} of all M ships and taking into account that $||Aw - Bv^{s} -c||^2 = \sum_{i \in \Mc}||A_iw_i - v^{s} -c_i||^2$.
	Since \eqref{ADMM-update-2} is identical to the NADMM update \eqref{NADMM}, the convergence of Algorithm \ref{sync-CCAS} is equivalent to the convergence of the NADMM for problem \eqref{prob2}.
	\begin{lemma}\label{ADMM_requirement}
		Consider the ADMM problem \eqref{prob2}. The following statements hold for all $i\in \Mc$:
		\begin{enumerate}[(i)]
			\item\label{lemma:lsc} $f_i$ is proper and lsc, and $g$ is proper, closed and convex.
			\item\label{lemma:surjective} $A$ is surjective.
			\item\label{lemma:L-smooth} $(A_i f_i)$ is $L_{(A_i f_i)}-\text{smooth}$, with constant $L_{(A_i f_i)}\in (0,+\infty)$. Hence $(Af)$ is $L-\text{smooth}$, with $L = \max_{i\in\Mc}\{L_{(A_i f_i)}\}$.
		\end{enumerate}			
	\end{lemma}
	\begin{proof}
		See Appendix \ref{lemma-proof}.
	\end{proof}
	Lemma \ref{ADMM_requirement} shows that the problem \eqref{prob2} satisfies the requirement for the NADMM formulation \cite[Asm. II]{themelis_20}.
	Therefore, due to the convergence of NADMM, we have the convergence of Algorithm \ref{sync-CCAS} as in Theorem \ref{proof-sync-CCAS}.
	\begin{theorem}[Convergence of the Algorithm \ref{sync-CCAS}]\label{proof-sync-CCAS}
		Assume that problem \eqref{prob2} has a feasible solution and the communication between ships is synchronous. Let $\left\{w^s_i,z_i^s,\xi^s\right\}$ be generated by Algorithm \ref{sync-CCAS} with $\lambda \in (0,2)$, $\beta > 2L$, and $L>0$. Then, the following holds for all $i \in \Mc$:
		\begin{enumerate}
			\item  The residual $(Aw^s_i-\xi^s-c_i)_{k\in \Nbb}$ vanish with $\min_{i\leq k}||(Aw^s_i-\xi^s-c_i||=o(1/\sqrt{k})$.
			\item $\left\{w^s_i,z_i^s,\xi^s\right\}$ converges to a stationary point $\left\{w^*_i,z_i^*,\xi^*\right\}$.
		\end{enumerate}
	\end{theorem}
	%
	%
	\section{Asynchronous Distributed Collaborative Collision Avoidance Algorithm}\label{sec:async-COLAV}
	\begin{figure}[!b]
		\centering
		\includegraphics[width=0.8\linewidth]{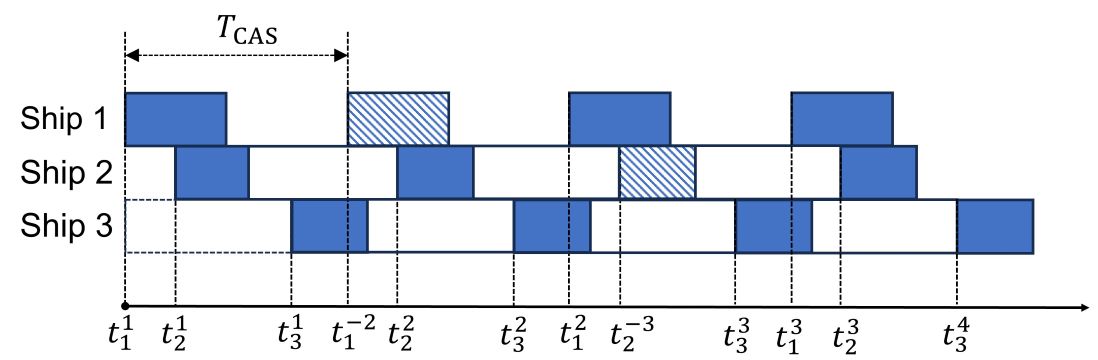}
		\caption{Example of the synchronous update scheme.}
		\label{example-sync}
	\end{figure}
	Although the Sync-CCAS algorithm can guarantee a stationary solution for problem \eqref{prob1}, it can be challenging to implement Sync-CCAS in real-time systems.
	Consider the following example with three ships with no coordinator (see Fig. \ref{example-sync}).
	Suppose three ships execute the Sync-CCAS algorithm after every time interval $T_\mathrm{CAS}$.
	Ship 1, 2, and 3 start the algorithm at time $t_1^1$, $t_2^1$, and $t_3^1$ respectively.
	After the duration of $T_\mathrm{CAS}$ from the first iteration, ship 1 is activated again at time $t_1^{-2}$, but cannot proceed to the second iteration due to missing information from ship 3.
	Therefore, ship 1 must skip update at time $t_1^{-2}$ and wait until the time $t_1^2$.
	As illustrated, it can take up to twice the interval time, i.e., $2T_\mathrm{CAS}$, to proceed with one interaction of Sync-CCAS.
	This delay time causes ships to respond slower to collision risks and harms their safety.
	This section presents an asynchronous version of the sync-CCAS algorithm, the Async-CCAS algorithm, to overcome the problem mentioned above.
	The Async-CCAS allows ships to update their decision without waiting for others, i.e. the asynchronous communication.
	To develop the Async-CCAS algorithm, we introduce the Async-NADMM.
	\subsection{Asynchronous Nonlinear Alternating Direction Method of Multipliers}
	\begin{algorithm}[!b]
		\caption{Async-NADMM} \label{R-ADMM-alg}
		\hspace*{\algorithmicindent} \textbf{Input:} Initiate the server with $v=v^0$, and clients with $z_i=z_i^0$, $w_i=w_i^0$, for $i\in \Mc$.
		\begin{algorithmic}[1]
			\FOR{ $s =1,..., s_{\max}$}
			\STATE \label{random} Choose client $i_s$ from $\Mc$ with equal probability.
			\STATE Client $i_s$ receive update of $v^s(t-d_{i_s})$ from coordinator with a delay $d_{i_s}$.
			\STATE [\textit{local update}] Client $i_s$ update local variables $w_{i_s}^{s+1}$, $z_{i_s}^{s+1}$ follow \eqref{w_i}, \eqref{z1}, and send the result to coordinator.
			\STATE [coordinator update] The server updates global variable $v^s$ using \eqref{v}.
			\STATE s:=s+1.
			\ENDFOR
		\end{algorithmic}
	\end{algorithm}
	We derive the Async-NADMM algorithm from the asynchronous randomized Douglas-Racford splitting (asynFedDR) algorithm that was proposed in \cite{tran_dinh_feddr_2021}.
	Consider the following optimization problem:
	\begin{subequations}\label{DR-prob}
		\begin{align}
			\min_{u_i,\tau \in \Rbb^p}  & \quad \frac{1}{M} \sum_{i=1}^{M} \varphi_i(u_i) +\phi(\tau), \\
			s.t. 				 & \quad u_i = \tau~~ \forall i \in \Mc.
		\end{align}
	\end{subequations}
	Here, $\varphi_i$ is a nonconvex and $L_{\varphi_i}$-smooth extended-real-valued functions; $\phi$ is a proper, closed and convex extended-real-valued function.
	Then, a Douglas-Rachford-iteration according to the asyncFedDR algorithm is as follows:
	\begin{align}\label{FedDR}
		\begin{split}
			\varsigma_i^{s+1} &= \varsigma_i^s+ \lambda(u_i^s-\tau^s), ~~\forall i \in \Mc, \\
			u_i^{s+1} &= \textbf{prox}_{\gamma\varphi_i} (\varsigma_i^{s+1}), ~~\forall i \in \Mc, \\
			\tau^{s+1}   &= \textbf{prox}_{\gamma \phi}\left[\frac{1}{M}\sum_{i=1}^{M}(2u_i^{s+1}-\varsigma_i^{s+1})\right]
		\end{split}
	\end{align}
	where $s$ is the iteration index.
	%
	%
	
	Based on the Async-FedDR, the Async-NADMM update for problem \eqref{prob2} is:
	\begin{subequations}\label{R-NADMM}
		\begin{align}
			z_i^{s+0.5} &= z_i^s -\beta(1-\lambda)(A_i w_i^s-v-c_i^s)~\forall i \in \Mc, \label{z0.5} \\
			w_i^{s+1} &= \arg \min_{w_i}\Bigl\{f_i(w_i)+\left\langle z_i^{s+0.5}, A_i w_i -v^{s} -c_i\right\rangle \nonumber\\
			&~~~~~~~~~~~~~~~~~~ +\frac{\beta}{2}||A_i w_i +v^{s} -c_i||^2\Bigr\} \label{w_i}\\
			\tilde{w}^{s+1} &=\frac{1}{M}\sum_{i=1}^{M}(A_i w_i^{s+1} -c_i), \\
			z_i^{s+1} &= z_i^{s+0.5} +\beta(A_i w^{s+1}_i-v^{s}-c_i)~\forall i \in \Mc, \label{z1} \\
			v^{s+1} &=  \arg \min_{v}\biggl\{g(v)+\Bigl\langle \frac{1}{M}\sum_{i=1}^{M} z_i^{s+1}, \tilde{w}^{s+1}-v \Bigr \rangle\nonumber\\
			&~~~~~~~~~~~~~~~~ +\frac{\beta}{2}||\tilde{w}^{s+1}-v||^2\biggr\}. \label{v}
		\end{align}
	\end{subequations}
	The algorithm of the Async-NADMM is presented in Algorithm \ref{R-ADMM-alg}.
	In this algorithm, a client can be seen as a controller that executes the algorithm, and the server can be one of the controllers that is designated to store the global variable.
	Additionally, if all controllers store their own version of the global variable and update its global variable's version whenever they receive an update from others, the update scheme is the same as that of Algorithm \ref{sync-CCAS}.
	The update scheme will be further discussed in Section \ref{sec:async-CCAS-B}.

	Similar to the asyncFedDR algorithm, at each iteration $s$ client $i_s$ receive a copy of global variable after a time delay $d_{i_s}$, i.e., $v^s(t_s-d_{i_s})$, with the current time $t_s$.
	Each client performs the updates asynchronously without waiting for the others to finish.
	Furthermore, 
	\begin{asst} \label{comm-prob}
		Assume that each client is activated at least once during $s_{\max}$ iterations of the Async-NADMM.
		Moreover, we assume that the time delays are bounded by $T_\mathrm{delay}<\infty$, i.e., $0 \leq d_{i_s}\leq T_\mathrm{delay}$ for all $i_s$.
	\end{asst}
	
	In order to show the convergence of Algorithm \ref{R-ADMM-alg}, we first show the universal equivalent between the Async-NADMM update \eqref{R-NADMM} and the asyncFedDR update \eqref{FedDR}.
	\begin{theorem}\label{ADMM equivalent}
		Let $(w_i,v,z_i)$ be generated by \eqref{R-NADMM} with relaxation $\lambda$ and penalty $\beta$ large enough so that every ADMM minimization subproblem has a solution. Let
		\begin{align}\label{drs-admm:var}
			\begin{cases}
				\varsigma_i^s &= A_i w_i^s -c_i - \frac{1}{\beta} z^s, \\
				u_i^s &= A_i w_i^s -c_i, \\
				\tau^s   &= v^s,
			\end{cases}
		\end{align}
		and 
		\begin{align*}
			\begin{cases}
				\varphi_i &= (A_if_i)(.+c_i),\\
				\phi &= g, \\
				\gamma &= \frac{1}{\beta}.
			\end{cases}
		\end{align*}
		Then we have the following relation:
		\begin{align}\label{admm-drs-update}
			\begin{cases}
				\varsigma_i^{s+1} &= \varsigma_i^s+ \lambda(u_i^s-\tau^s)~~\forall i \in \Mc, \\
				u_i^{s+1} &= \textbf{prox}_{\gamma\varphi_i} (\varsigma_i^{s+1}) ~~\forall i \in \Mc, \\
				\tau^{s+1}   &= \textbf{prox}_{\gamma \phi}\left\{\frac{1}{M}\sum_{i=1}^{M}(2u_i^{s+1}-\varsigma_i^{s+1})\right\}.
			\end{cases}
		\end{align}
		Moreover, if $A_i$ is surjective for all $i \in \Mc$, and $\beta> L$, then it holds that
		\begin{align}\label{dre}
			V^s_{\gamma} = \Lc_\beta^s,
		\end{align}
		where
		\begin{align*}
			V^s_\gamma(\tau^s) &= \phi(\tau^s)+\frac{1}{M}\sum_{i=1}^{M}\Bigl[\varphi_i(u_i^s) \\
			&~+\langle\nabla\varphi_i(u_i^s),\tau^s-u_i^s\rangle+\frac{1}{2\gamma}||\tau^s-u_i^s||^2\Bigr],\\
			\Lc_\beta^s &=g(v^s) + \frac{1}{M} \sum_{i=1}^{M} \Bigl[ f_i(w_i^s)  + \langle z_i^s,A_i w^{s}_i-v^{s}-c_i\rangle \\
			&~~~~~~~~~~~~~~+\frac{\beta}{2}||A_i w^{s}_i-v^{s}-c_i||^2\Bigr].
		\end{align*}
	\end{theorem}
	\begin{proof}
		See appendix \ref{R-ADMM-proof}
	\end{proof}
	\begin{algorithm}[!t]
		\caption{Async-CCAS} \label{async-CCAS}
		\begin{algorithmic}[1]
			\FOR{ $s = 1,...,s_{\max}$}
			\FORALL{$i \in \Mc$ each ship computes in parallel}	
			\STATE Receive $\hat{\xi}_j^{s}$ from neighboring ships, $j \in \Mc \backslash \{ i\}$
			\IF{$\hat{\xi}_j^{s}$ is unavailable}
			\STATE $\hat{\xi}_j^{s} = \hat{\xi}_j^{s-1}$
			\ENDIF
			\STATE Update the global variable $\xi^{s+1}$ using \eqref{update_global}
			\STATE Update local variables $z^{s+1}_{i}$, $\tilde{u}_i^{s+1}$, and  $\tilde{p}_{i}^{s+1}$ using \eqref{update_z1}--\eqref{update_global_local} \label{async:updateADMM}
			\STATE $\hat{\xi}_j^{s}=\tilde{R}^{-1}_i\tilde{p}_i^{s} + \bar{\eta}_{i}^\mathrm{WP}$ \label{async:transform}
			\STATE Transmit data $\hat{\xi}_i^{s} $ to all ship $j \in \Mc \backslash \{ i\}$. \label{async:transmit}
			\STATE $s := s+1 $.
			\ENDFOR
			\ENDFOR
		\end{algorithmic}
	\end{algorithm}

	Theorem \ref{ADMM equivalent} allows us to employ the convergence analysis of the asyncFedDR to the Async-NADMM.
	\begin{theorem}\label{ADMM-convergence}
		Suppose that Problem \eqref{prob2} has a feasible solution with $F^* := \inf F(w,v) >-\infty$ and Assumptions \ref{comm-prob} holds. Let $(w_i,v,z_i)$ be generated by \eqref{R-NADMM} with relaxation $\lambda$, penalty $\beta > L$. Then the following hold:
		\begin{enumerate}[(i)]
			\item \label{ADMM-conv1} $(w^s,v^s)$ converges to $(w^*,v^*)$ such that
			\begin{align*}
				\Ebb \left[\big|\big|\nabla F(w^*,v^*)\big|\big|^2\right]<\epsilon^2
			\end{align*}
			where the expectation is taken with respect to all random variables and communication delays in the Async-ADMM algorithm, and $\epsilon$ is an arbitrary positive number.
			\item \label{ADMM-conv2} The residual $A_i w^{s}_i-v^{s}-c_i$ is mean-square convergent to $0$ with the rate $\min_{p\leq s}\Ebb[||A_i w^{p}_i-v^{p}-c_i||^2] = o(1/s)$.
			
		\end{enumerate}
	\end{theorem}
	\begin{proof}
		Theorem \ref{ADMM-convergence}\ref{ADMM-conv1} is a direct application of Theorem 4.1 from \cite{tran_dinh_feddr_2021}.
		
		Theorem \ref{ADMM-convergence}\ref{ADMM-conv2}. Lemma B.3 and the proof of Theorem 4.1 in \cite{tran_dinh_feddr_2021} imply that:
		\begin{align}\label{ADMM-conv-proof1}
			\sum_{s\in \Rbb} \sum_{i=1}^{M}\Ebb\left(||\tau^s-u_i^s||^2\right) \leq D [F(w^0,v^0)-F^*],
		\end{align}
		where $D$ is a positive constant defined as in \cite[Lem. B.3]{tran_dinh_feddr_2021}.
		Using \eqref{drs-admm:var}, then \eqref{ADMM-conv-proof1} is equivalent to:
		\begin{align}
			\sum_{s\in \Rbb}\sum_{i=1}^{M} \Ebb\left(||A_i w^{s}_i-v^{s}-c_i||^2\right) \leq D [F(w^0,v^0)-F^*].
		\end{align}
		Since $F^* > -\infty$ and $F(w^0,v^0) \in \Rbb$, then $(A_i w^{s}_i-v^{s}-c_i)_{s\in \Nbb}$ is mean-square convergent to $0$ with the claimed rate.
		
	\end{proof}

	In Theorem \ref{ADMM-convergence}\ref{ADMM-conv1}, random variables exist due to step \ref{random} of Algorithm \ref{R-ADMM-alg}, where we randomly chose a client $i_s$ to perform the updates.
	\subsection{Asynchronous distributed collision avoidance algorithm}\label{sec:async-CCAS-B}
	\begin{figure}[!b]
		\centering
		\includegraphics[width=0.8\linewidth]{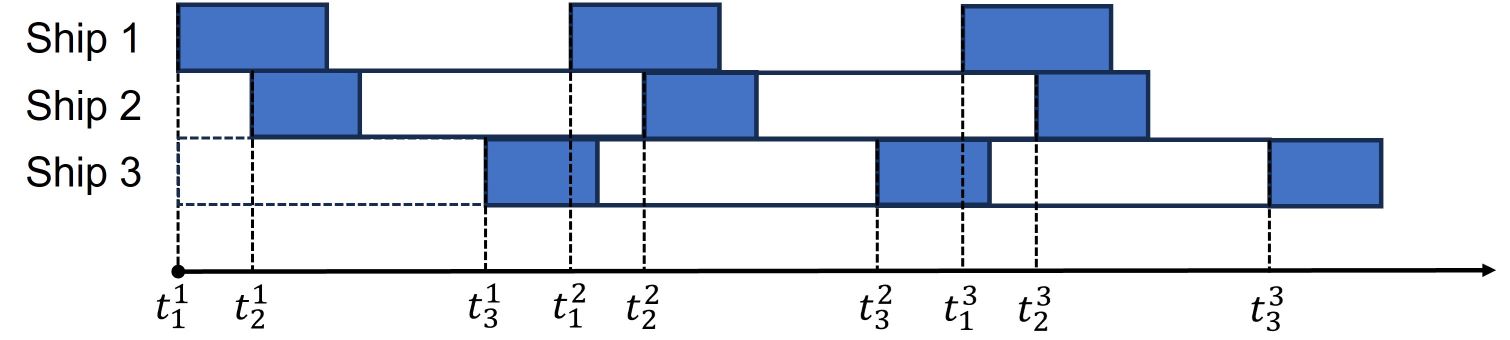}
		\caption{Example of the asynchronous update scheme.}
		\label{example}
	\end{figure}
	We introduce the asynchronous update scheme as in Algorithm \ref{async-CCAS}.
	Unlike Algorithm \ref{sync-CCAS}, each ship does not need to wait for others to finish their update to proceed to the next iteration.
	Due to this asynchronous scheme, the global variables that each ship uses may differ.
	This can be illustrated in the following example (see Fig. \ref{example}).
	Suppose that we have three ships with different starting times. 
	Besides, each ship needs a different time to perform the updates (the blue area in the timeline).
	At time $t= t^1_1$ and $t=t^1_2$, ship 1 and 2 perform global updates using $\left[\hat{\xi}_1^{0},\hat{\xi}_2^{0},\hat{\xi}_3^{0}\right]$.
	Then at time $t= t^1_3$, ship 3 performs global update using $\left[\hat{\xi}_1^{1},\hat{\xi}_2^{1},\hat{\xi}_3^{0}\right]$.
	At the second iteration ($s=2$), we also see that ship 1 performs the global update with $\left[\hat{\xi}_1^{1},\hat{\xi}_2^{1},\hat{\xi}_3^{0}\right]$.
	However, ship 2, at $t= t^2_2$,  performs update with $\left[\hat{\xi}_1^{1},\hat{\xi}_2^{1},\hat{\xi}_3^{1}\right]$.
	As can be seen, each ship uses a different set of $\hat{\xi}_j^{s}$ when performing the global update, and this situation is considered by AsyncFedDR.
	Therefore, the Async-ADMM, an application of AsyncFedDR, was utilized to construct Algorithm \ref{async-CCAS}.

	Given that each ship performs the updates after a constant time period, $\Delta T_\mathrm{CAS}$, and assuming that the calculation time is less than $\Delta T_\mathrm{CAS}$, the maximum delay time of the global variable $T_\mathrm{delay}$ is bounded by $\Delta T_\mathrm{CAS}$ plus the data transmission time $T_\mathrm{trans}$, i.e., $T_\mathrm{delay}<\Delta T_\mathrm{CAS}+T_\mathrm{trans}$.
	If we further assume that the data transmission time is bounded, then Algorithm \ref{async-CCAS} satisfies Assumption \ref{comm-prob}.
	%
	%
	Therefore, the convergence of Algorithm \ref{async-CCAS} is guaranteed following Theorem \ref{ADMM-convergence}.
	\begin{table}[!b]
		\centering
		\caption{Control parameters}
		{\renewcommand{\arraystretch}{1.5}
			\begin{tabular}{|c|c|c|c|}
				\hline
				Parameter & Unit & Simulation & Field\\
				\hline
				$y_{\max}$& m    & 60         & 10\\
				\hline
				$\chi^{\max}$& rad    & $\pi/6$  & $\pi/6$\\
				\hline
				$K_y$& --   & $10^{-2}$  & $2\times10^{-2}$\\
				\hline
				$K_u$& --   & $2\times10^{-2}$  & $4\times10^{-2}$\\
				\hline
				$\beta$& --   & $3\times10^{-4}$  & $5\times10^{-4}$\\
				\hline
				$\Delta T$ & s & 20 & 5 \\
				\hline
		\end{tabular}}
		\label{tab:param}
	\end{table}

	\section{Experimental results}\label{sec:sim}
	This section presents the evaluation of the performance of the proposed CCAS algorithms in both simulation and field experiments, where we consider that the ship sails along inland waterways. 
	Therefore, the lateral control command $u^y_{i,i}$ is bounded by the width of the waterway. Although this, in practice, will depend on the lateral position of the ship, we represent this here with a constant symmetric constraint $|u^y_{i,i}|\leq y_{\max}$.
	The predicted control signal for neighboring ships is also bounded by the maximum course angle change that the ship can make during the update interval $\Delta T$, i.e., $0\leq\ix{\chi}{j}^d(k) \leq\Delta\chi^{\max}$.
	The control parameters are presented in Table \ref{tab:param}.
	Moreover, the Casadi toolbox \cite{Andersson2019} with the Interior Point Optimizer - $ipopt$ \cite{wachter_implementation_2004} is used to solve the local MPC problem \eqref{update_local}.
	%
	%
	
	%
	%
	%
	\begin{figure}[!t]
		\centering
		\includegraphics[width=0.6\linewidth]{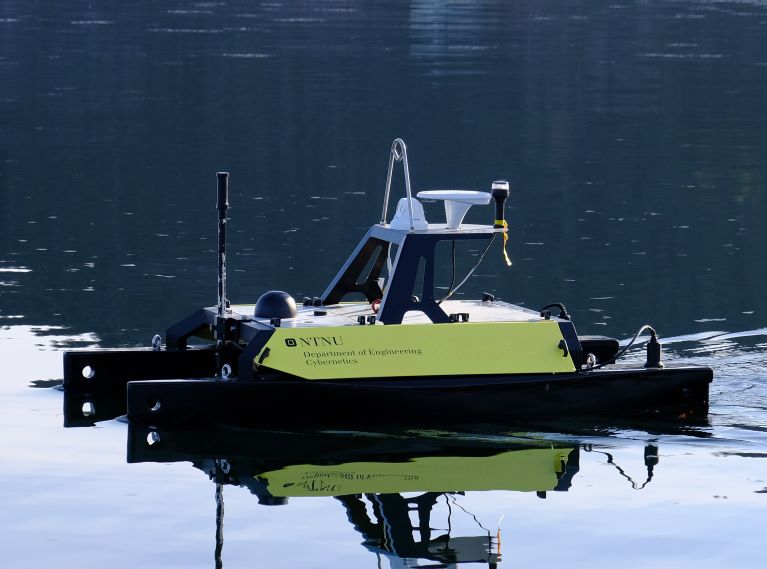}
		\caption{The NTNU Fish Otter.}
		\label{fig:otter}
	\end{figure}
	\subsection{Performance evaluation}
	We use two different ship models for simulation and field experiments.
	In field experiments, ships with dimensions of $2\text{m}\times1.08\text{m}$ are used.
	In simulation experiments, a ship model has a length of $51.5\text{m}$, and a beam of $8.6\text{m}$ is used.
	Due to the significance of the size and shape of the two models, we adopt two methods to evaluate the safety performance of the CCAS algorithms.
	
	In the field experiments, we define a threshold distance of $5\text{m}$ for the safety distance of the ship.
	Their navigation is safe if the distance $d_{ij}$ between ship $i$ and $j$ is greater or equal to the threshold.
	
	Due to the ship's dimensions, it is not practical to use the distance (from the center of mass) between two ships in simulation experiments to evaluate the safety of navigation.
	For example, the safety distance between two ships in a head-on situation can be as small as $16\text{m}$.
	%
	For this reason, we define a rectangular safety domain around the ship so that if there is no other ship within this domain, then the navigation is safe.
	An index $\epsilon_i$,  with $\epsilon_i\leq0$, indicates that there is another ship in the safety domain, is defined as follows: 
	\begin{align*}
		\epsilon_{i}&= \min_{j\in \Mc \backslash \{ i \} }\left\{\max\{|\ix{x}{i} - \ix{x}{j}|-d^x_i,|\ix{y}{i} - \ix{y}{j}|-d^y_i\}\right\},
	\end{align*}
	where $2d^x_i$ and $2d^y_i$ are the length and width of the rectangle safety area, respectively.
	In this simulation, $d^x_i=51.5 \ \text{m}$ and $d^y_i=8.6 \ \text{m}$.
	
	Besides the safety performance, the computational performance is evaluated based on the computational time that the CCAS algorithms need to perform an update.
	The computational time is acceptable if its maximum is less than the update interval $\Delta T$ (see Table \ref{tab:param}).
	%
	\subsection{Simulation experiments}
	\begin{table*}
		\begin{center}
			\begin{threeparttable}
				\caption{Computational time (in second) needed to perform an update in simulation experiments}
				\label{table:sync-ccas}
				
				{\renewcommand{\arraystretch}{1.25}
					\begin{tabular}{ |c |c|c|c|c|c|c|c| } 
						\hline
						\multicolumn{2}{|c|}{Scenario} & Ship 1 & Ship 2 & Ship 3 & Ship 4 & Ship 5 & Ship 6\\
						\hline
						\multirow{2}{10em}{4 ships, Sync-CCAS, $s_\mathrm{max}=2$ (Fig. \ref{fig:cross41})}& Max & 3.3091 & 2.9445 & 2.6709 & 3.43 & - & -\\
						\cline{2-8}
						& Mean & 2.2592 & 1.6333 & 1.1366 & 1.9705 & - & -\\
						\cline{2-8}
						\cline{2-8}
						\hline
						\multirow{2}{10em}{4 ships, Async-CCAS, $s_\mathrm{max}=2$}& Max & 2.6784 & 2.9268 & 2.7204 & 2.8025 & - & -\\
						\cline{2-8}
						& Mean & 1.8959 & 1.5590 & 1.2678 & 1.7648 & - & -\\
						\cline{2-8}
						\cline{2-8}
						\hline
						\multirow{2}{10em}{4 ships, Async-CCAS, $s_\mathrm{max}=5$}& Max & 2.7238 & 2.6739 & 2.7102 & 2.9427 & - & -\\
						\cline{2-8}
						& Mean & 1.7888 & 1.5838 & 1.3552 & 1.6359 & - & -\\
						\cline{2-8}
						\cline{2-8}
						\hline
						\multirow{2}{10em}{4 ships, Async-CCAS, $s_\mathrm{max}=10$}& Max & 3.0268 & 3.1482 & 2.7383 & 3.093 & - & -\\
						\cline{2-8}
						& Mean & 1.7934 & 1.5658 & 1.3912 & 1.9140 & - & -\\
						\cline{2-8}
						\cline{2-8}
						\hline
						\multirow{2}{10em}{6 ships, Sync-CCAS, $s_\mathrm{max}=2$ (Fig. \ref{fig:cross61})} & Max & 6.5665 & 4.289 & 5.1902 & 5.6937 & 4.8888 & 5.1652\\
						\cline{2-8}
						& Mean & 3.6829 & 2.52 & 2.446 & 3.4739 & 3.0647 & 3.2266\\
						\cline{2-8}
						\cline{2-8}
						\hline
						\multirow{2}{10em}{6 ships, Async-CCAS, $s_\mathrm{max}=2$ (Fig. \ref{fig:cross62})} & Max & 4.9001 & 4.1184 & 7.6309 & 5.5773 & 4.0894 & 4.0667\\
						\cline{2-8}
						& Mean & 2.7975 & 1.9842 & 2.5506 & 2.6965 & 2.8335 & 2.5194\\
						\cline{2-8}
						\cline{2-8}
						\hline
						\multirow{2}{10em}{6 ships, Async-CCAS, $s_\mathrm{max}=2$, data loss $5\%$} & Max & 5.4527 & 5.7382 & 4.6384 & 5.6987 & 5.8374 & 6.7083\\
						\cline{2-8}
						& Mean & 3.5032 & 2.5749 & 2.5739 & 3.5062 & 3.2180 & 3.3332\\
						\cline{2-8}
						\cline{2-8}
						\hline
				\end{tabular}}
				
				\begin{tablenotes}
					\small
					\item $s_{\max}$: maximum iteration of CCAS algorithms.
				\end{tablenotes}
			\end{threeparttable}
		\end{center}
	\end{table*}
	\begin{figure*}[!t]
		\centering
		\subfloat[\label{fig:cross41_1}]{
			\centering
			\includegraphics[width=0.25\linewidth]{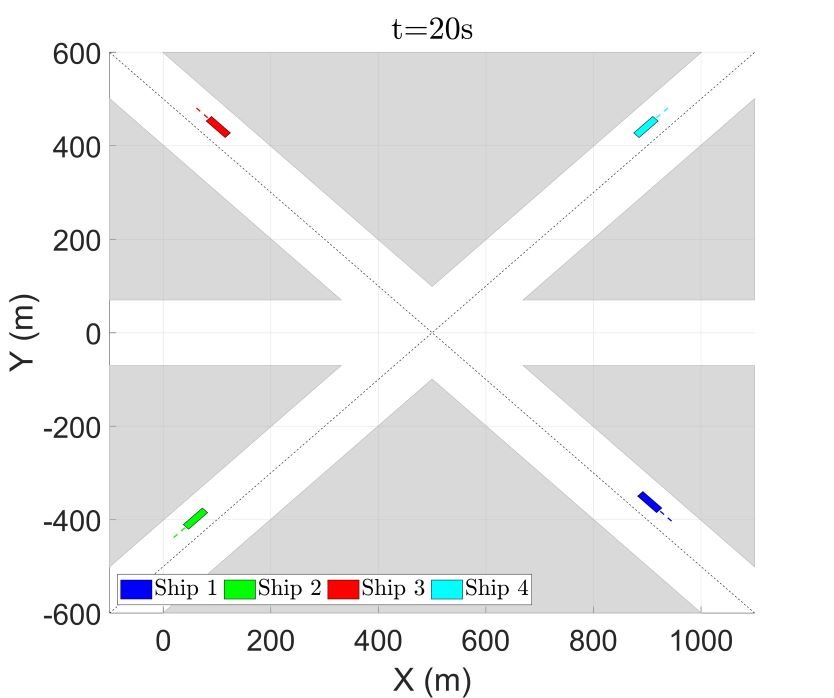}}
		\centering
		\subfloat[\label{fig:cross41_2}]{
			\centering
			\includegraphics[width=0.25\linewidth]{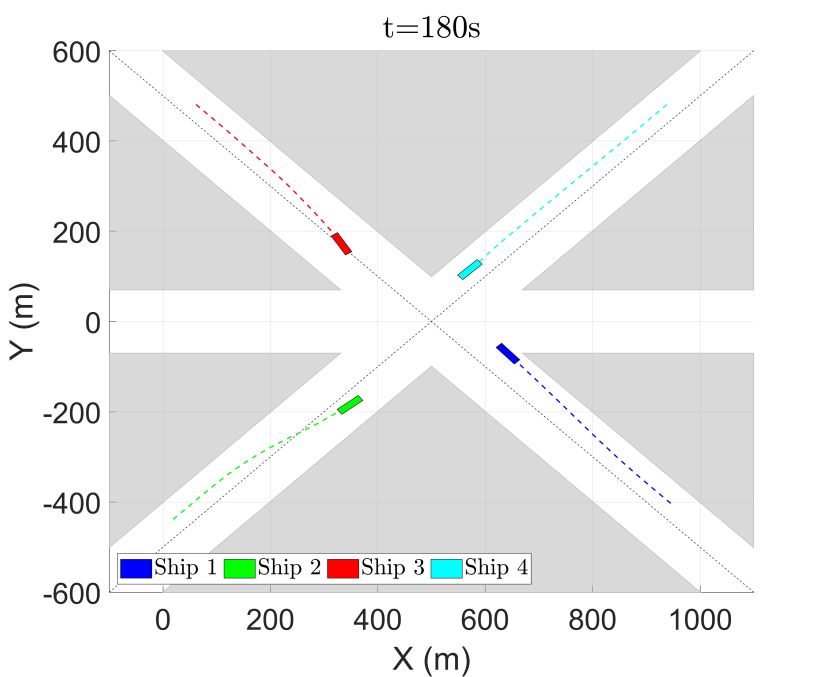}}
		\centering
		\subfloat[\label{fig:cross41_3}]{
			\centering
			\includegraphics[width=0.25\linewidth]{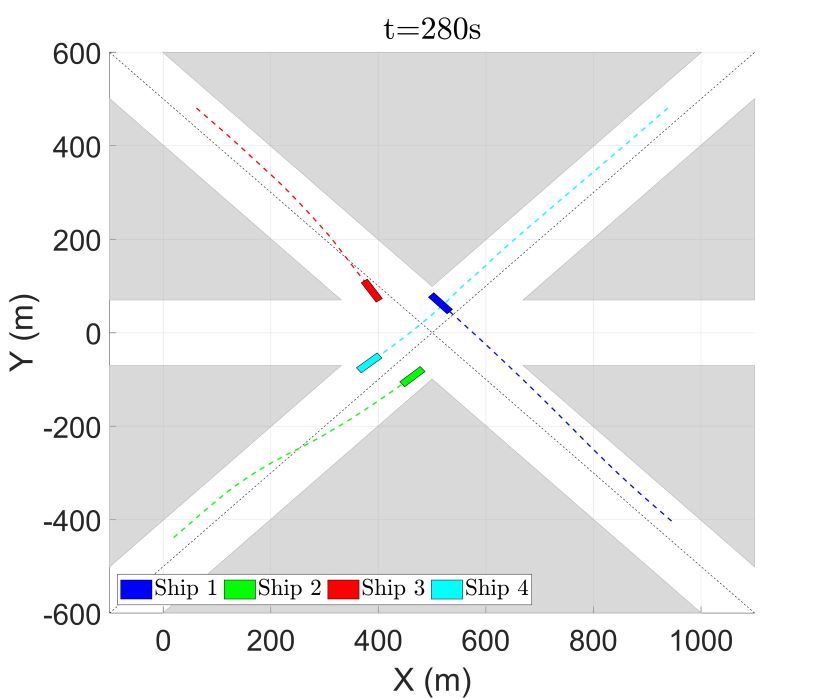}}
		\hfill
		\subfloat[\label{fig:cross41_4}]{
			\centering
			\includegraphics[width=0.25\linewidth]{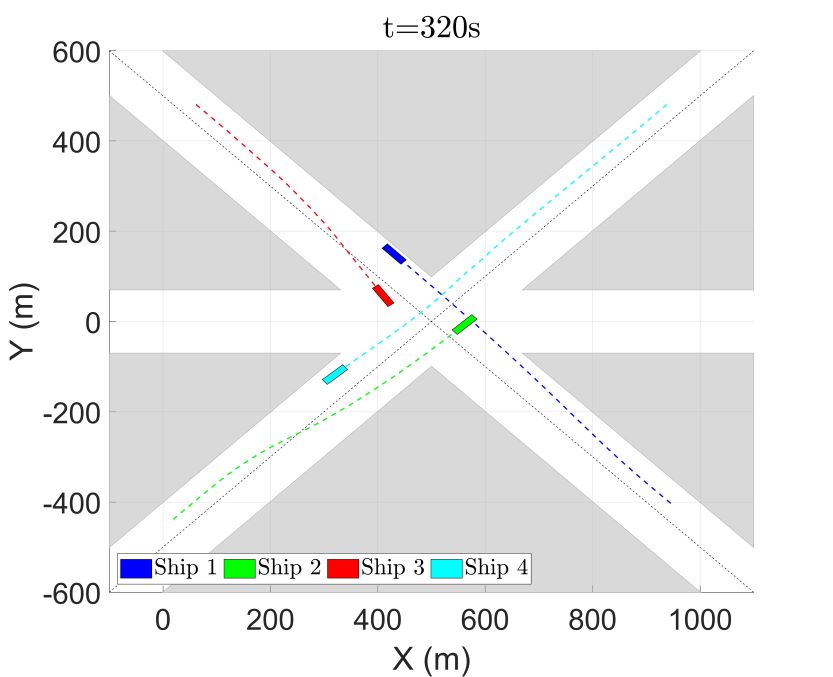}}
		\centering
		\subfloat[Safety indices of ships\label{fig:cross41_5}]{
			\centering
			\includegraphics[width=0.35\linewidth]{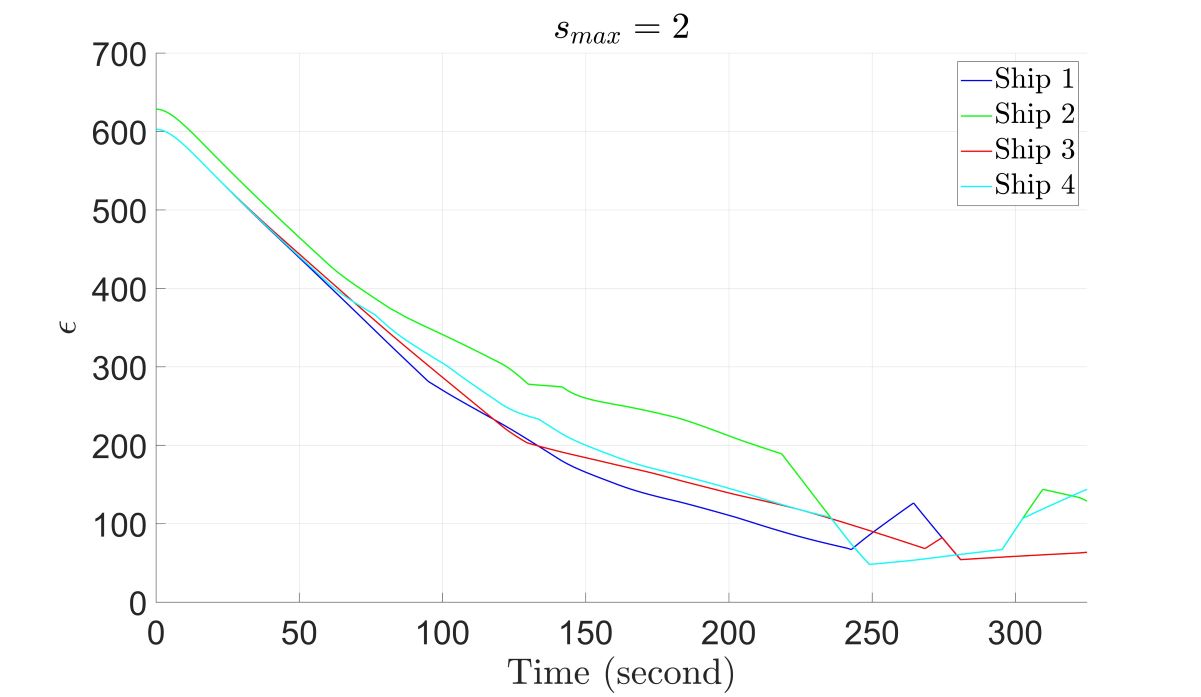}}
		\caption{Simulation experiments (Synchronous communication): Intersection crossing between 4 ships. The guiding lines are depicted in the black dashed lines.}
		\label{fig:cross41}
	\end{figure*}
	\begin{figure*}[!t]
		\centering
		\subfloat[\label{fig:cross412}]{
			\centering
			\includegraphics[width=0.25\linewidth]{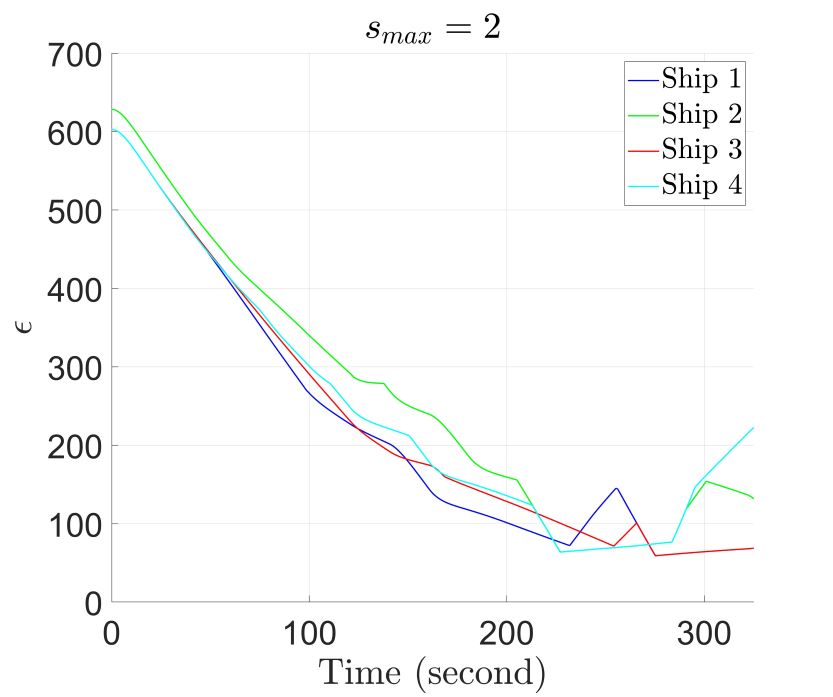}}
		\centering
		\subfloat[\label{fig:cross415}]{
			\centering
			\includegraphics[width=0.25\linewidth]{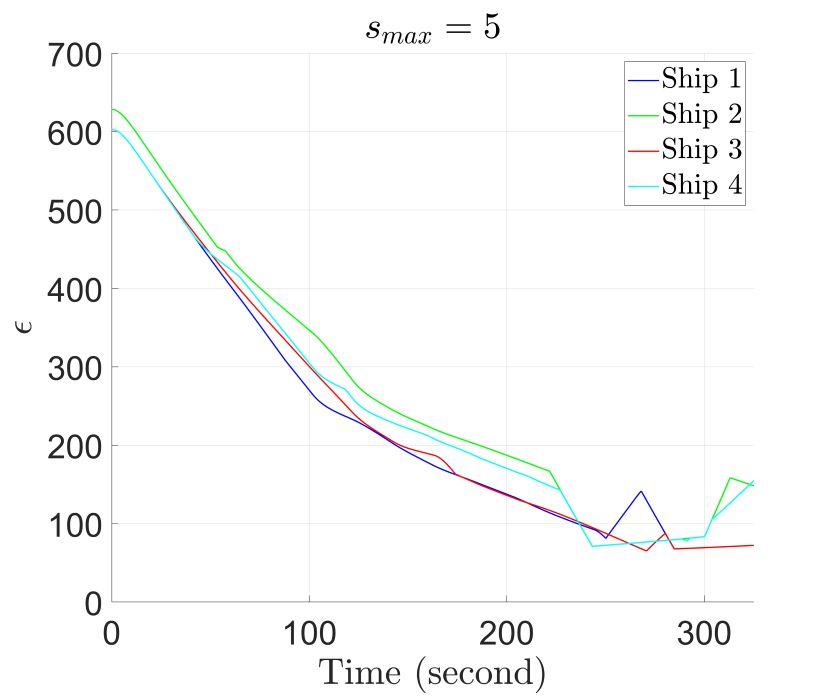}}
		\centering
		\subfloat[\label{fig:cross4110_3}]{
			\centering
			\includegraphics[width=0.25\linewidth]{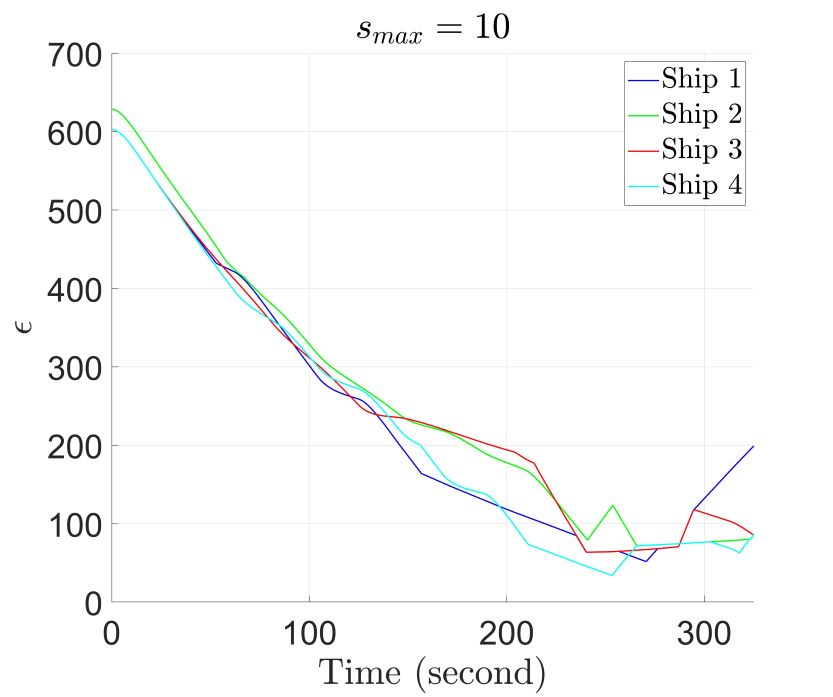}}
		\caption{Safety indices of 4 ships scenarios with different values of maximum iteration of CCAS algorithm $s_{\max}$.}
		\label{fig:smax}
	\end{figure*}
	\begin{figure*}[!t]
		\centering
		\subfloat[\label{fig:cross61_1}]{
			\centering
			\includegraphics[width=0.25\linewidth]{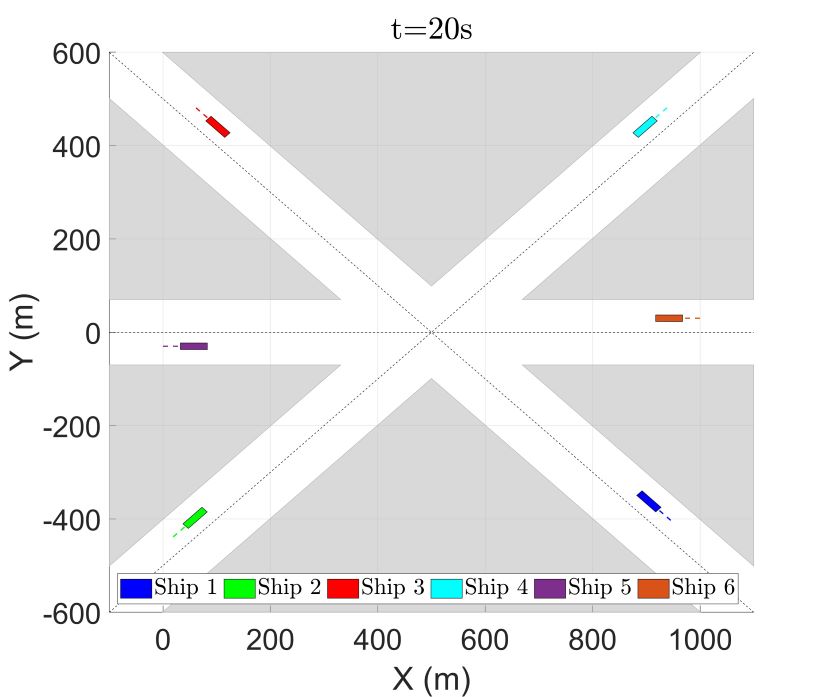}}
		\centering
		\subfloat[\label{fig:cross61_2}]{
			\centering
			\includegraphics[width=0.25\linewidth]{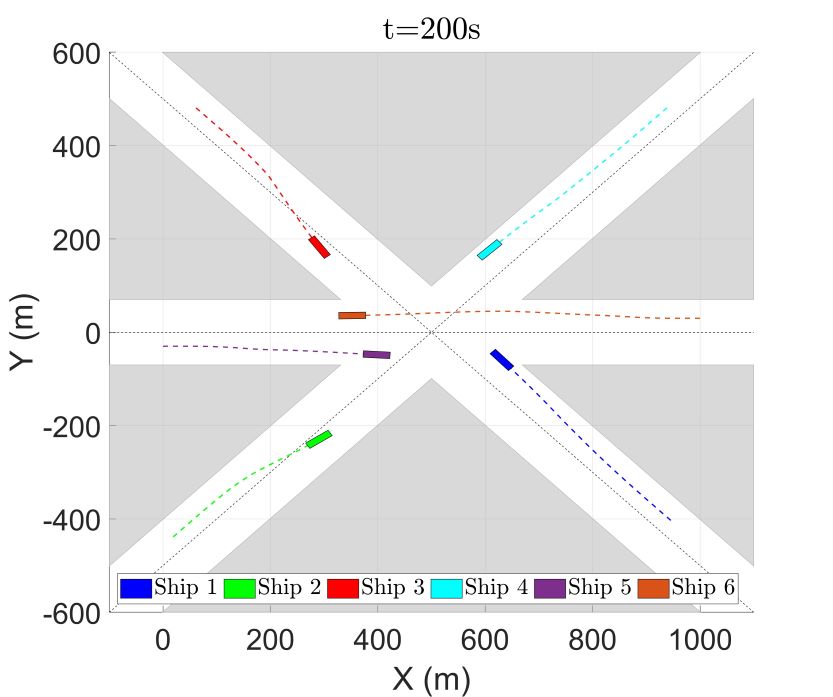}}
		\centering
		\subfloat[\label{fig:cross61_3}]{
			\centering
			\includegraphics[width=0.25\linewidth]{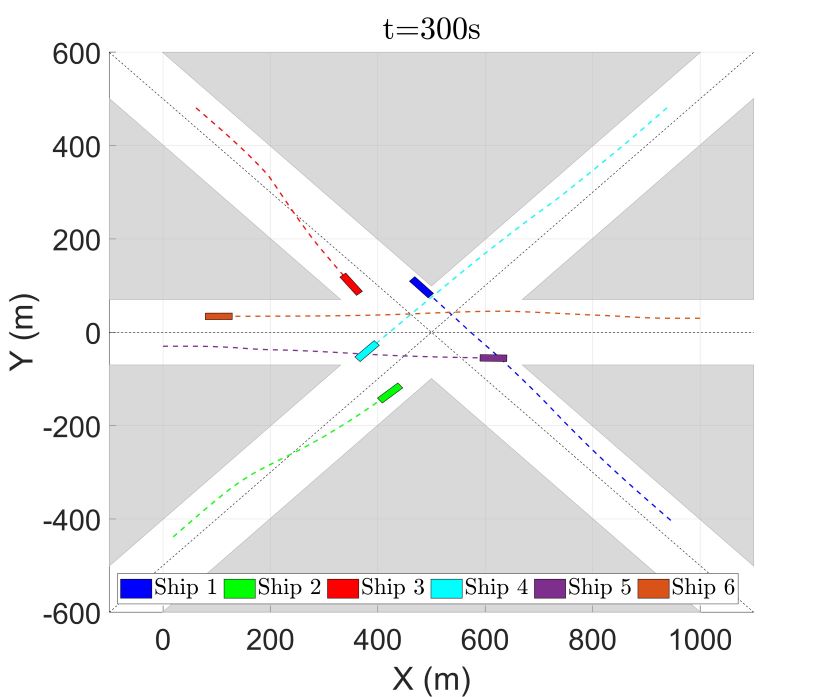}}
		\hfill
		\subfloat[\label{fig:cross61_4}]{
			\centering
			\includegraphics[width=0.25\linewidth]{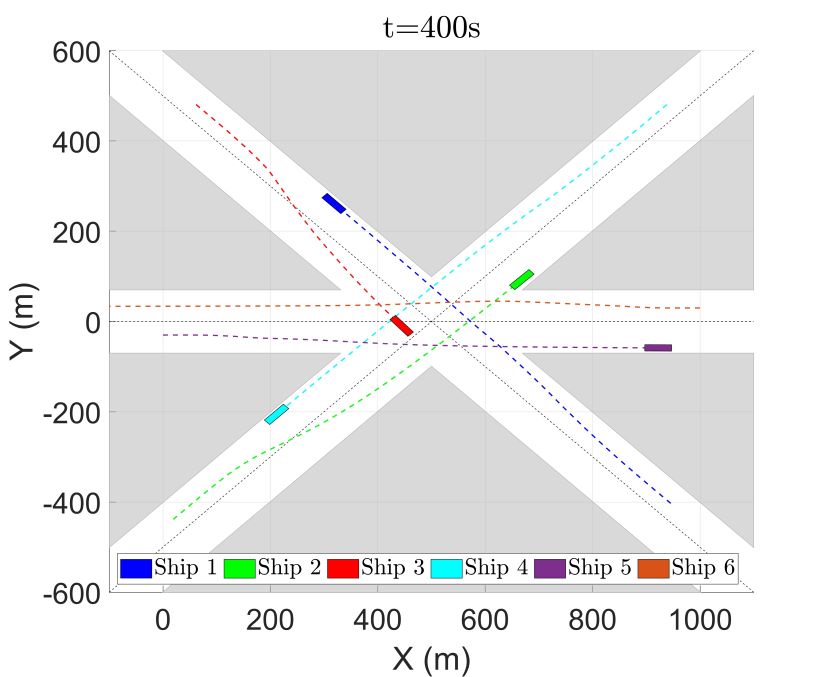}}
		\centering
		\subfloat[Safety indices of ships\label{fig:cross61_5}]{
			\centering
			\includegraphics[width=0.35\linewidth]{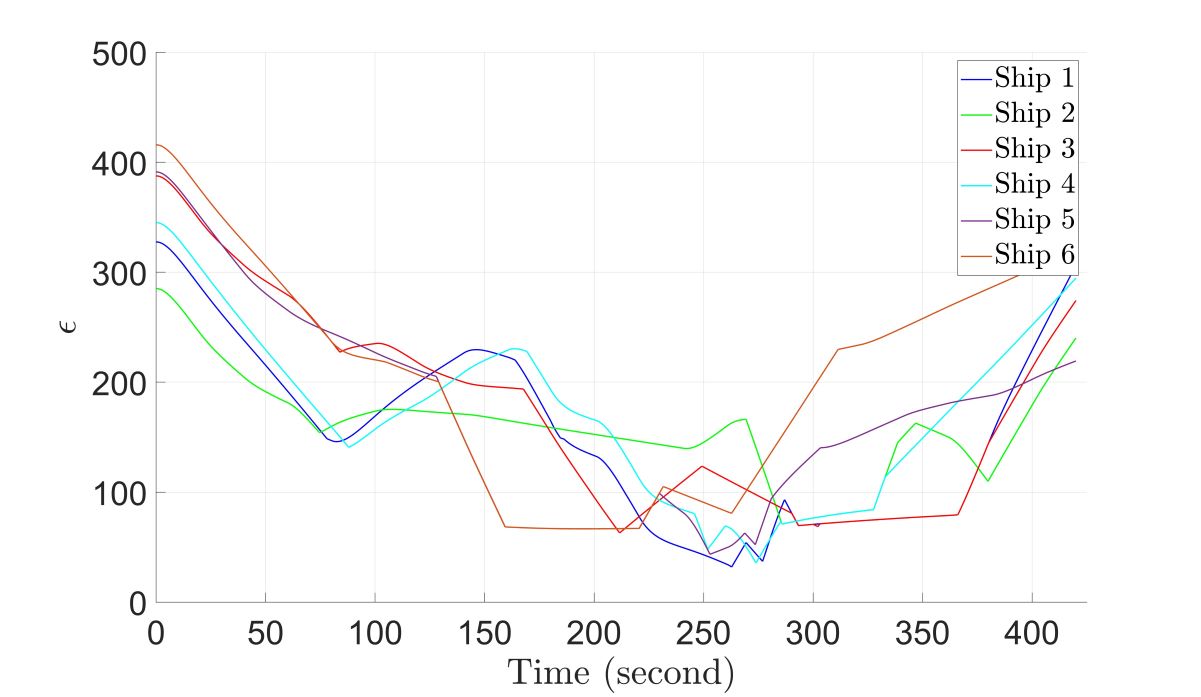}}
		\caption{Simulation experiments (Synchronous communication): Intersection crossing between 6 ships. The guiding lines are depicted in the black dashed lines.}
		\label{fig:cross61}
	\end{figure*}
	\begin{figure*}[!t]
		\centering
		\subfloat[\label{fig:cross62_1}]{
			\centering
			\includegraphics[width=0.25\linewidth]{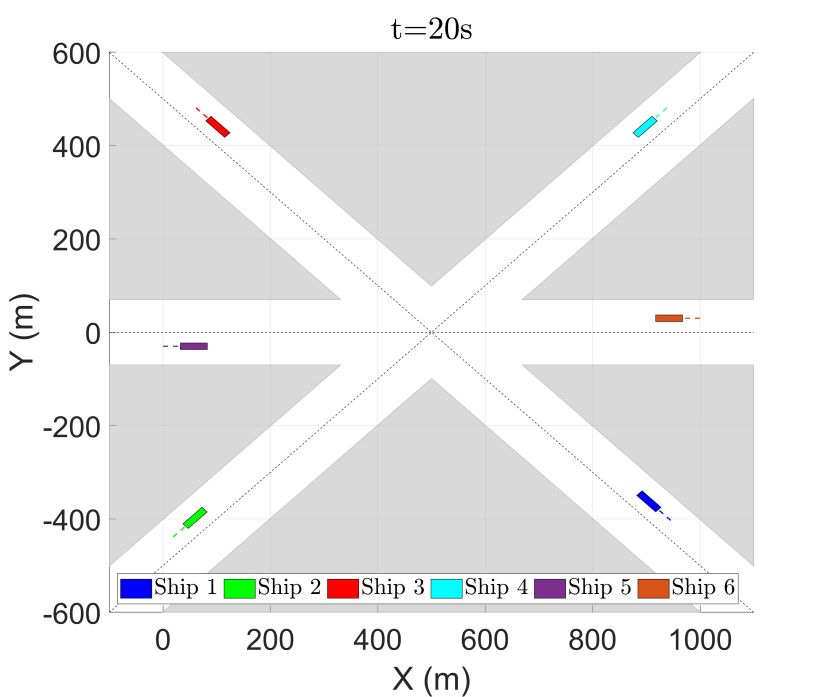}}
		\centering
		\subfloat[\label{fig:cross62_2}]{
			\centering
			\includegraphics[width=0.25\linewidth]{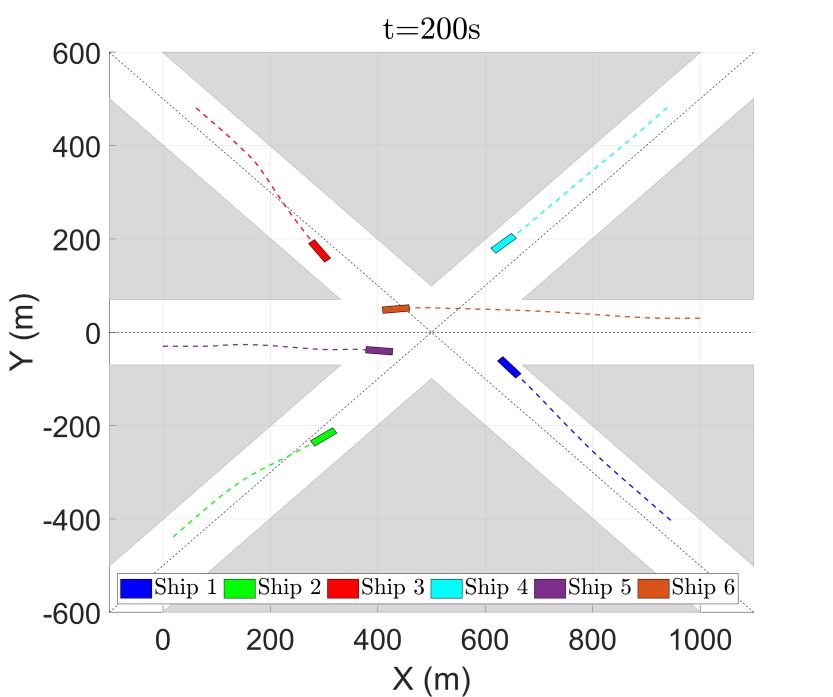}}
		\centering
		\subfloat[\label{fig:cross62_3}]{
			\centering
			\includegraphics[width=0.25\linewidth]{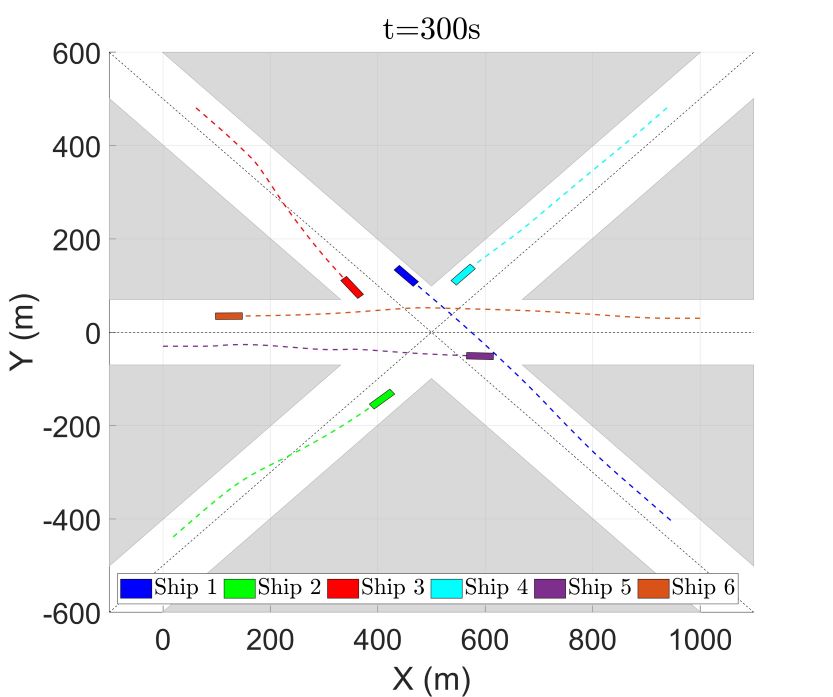}}
		\hfill
		\subfloat[\label{fig:cross62_4}]{
			\centering
			\includegraphics[width=0.25\linewidth]{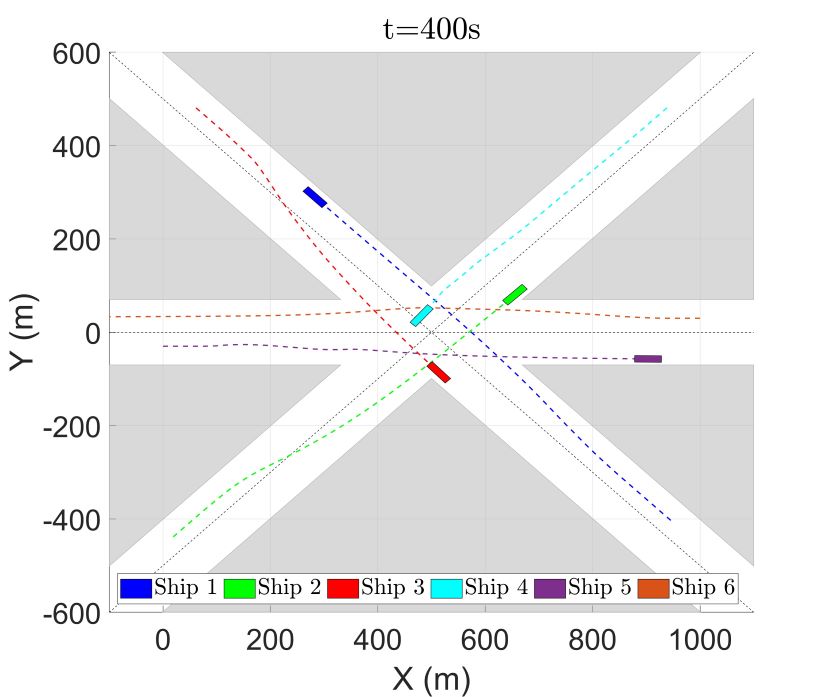}}
		\centering
		\subfloat[Safety indices of ships\label{fig:cross62_5}]{
			\centering
			\includegraphics[width=0.35\linewidth]{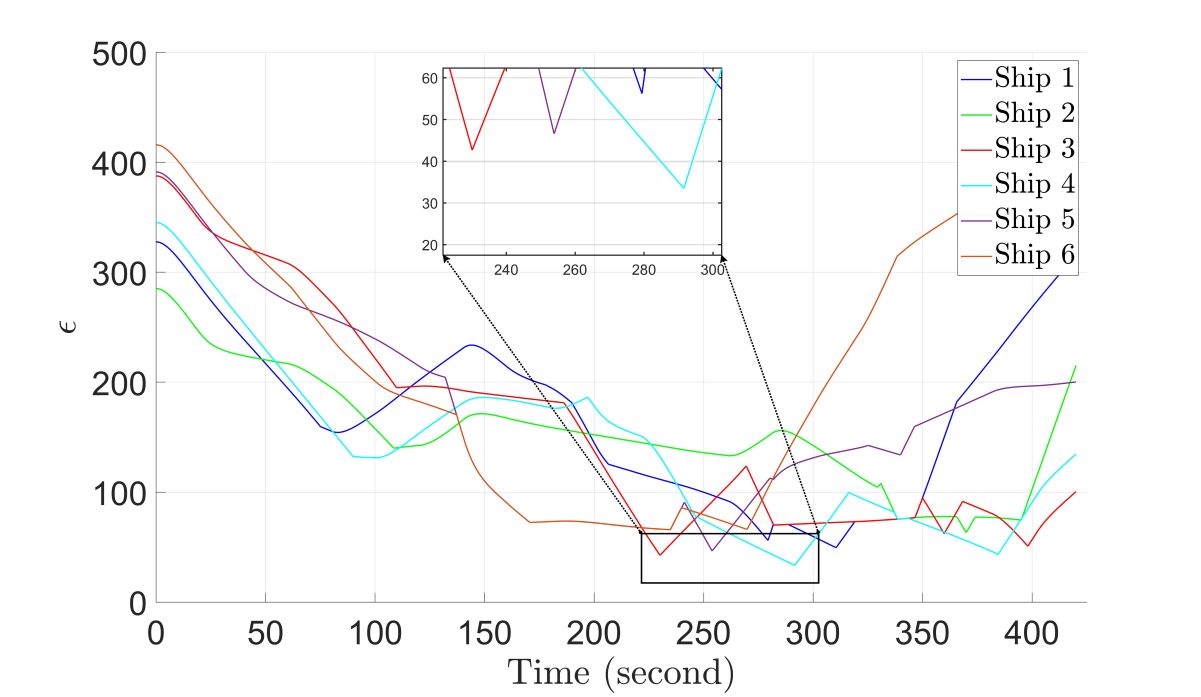}}
		\caption{Simulation experiments (Asynchronous communication): Intersection crossing between 6 ships. The guiding lines are depicted in the black dashed lines.}
		\label{fig:cross62}
	\end{figure*}
	\begin{figure}
		\centering
		\includegraphics[width=0.7\linewidth]{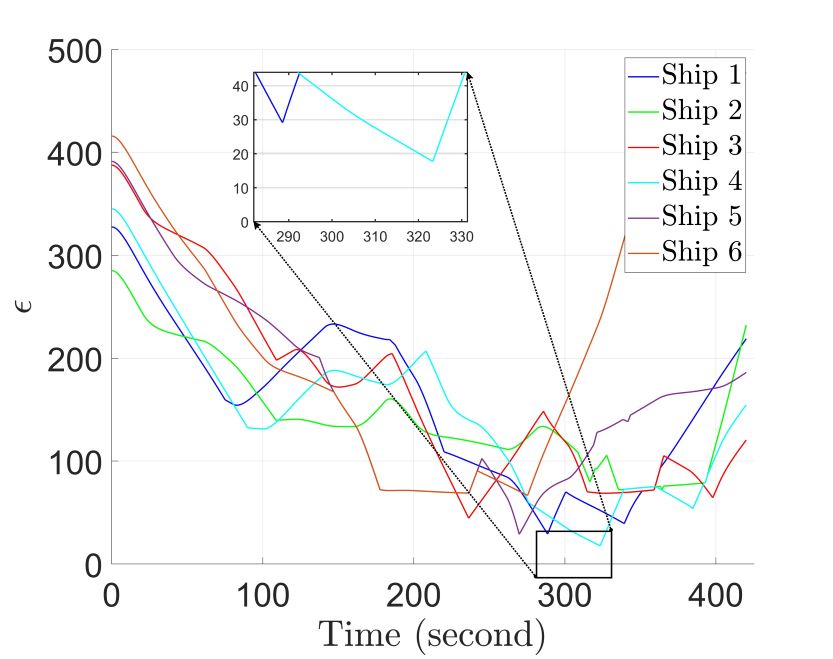}
		\caption{Safety indices in case of $5\%$ data loss.}
		\label{fig:loss5}
	\end{figure}
	The simulation experiments are done in Matlab 2024a running on a PC with an Intel(R) Core(TM) i7-11850H and 32GB of RAM.
	The CCAS algorithm is called every 20 seconds.
	%
	%
	%
	
	The purposes of the simulation experiments are to illustrate the scalability of the Sync-CCAS algorithm and compare it to the Async-CCAS algorithm.
	Therefore, only the scenarios of 4 and 6 ships are presented.
	We also present the computational time with each ship's maximum and mean time in Table \ref{table:sync-ccas}.
	
	In the simulation environment, we use a shared variable to store a copy of all the decisions of the ships, i.e., $\xi_i$.
	Each ship does not receive updates directly from $\xi_i$, but from the copy that is stored in the share variable.
	Depending on the time the shared variable is updated, we can simulate synchronous or asynchronous communication.
	To simulate synchronous communication, we update variables $\xi_i$ of all $M$ ships simultaneously after they have all finish their individual update.
	This way guarantees that all ships receive the same information for each iteration update.
	On the contrary, in asynchronous experiments, we update the shared variable every time a ship $i$ has updated its decision $\xi_i$.
	Due to multiple updates of the shared variable in one iteration, the ship that performs updates later can receive different inputs than the former.
	Although there is no actual delay in communication between ships in the simulation environment, thanks to the difference in inputs each ship receives, we can simulate asynchronous communication.  
	
	In the 4-ships scenario with Sync-CCAS algorithm (Fig. \ref{fig:cross41}), ships 1 and 4 have the stand-on priority but need to negotiate with each other to avoid a collision.
	On the other hand, ship 3 must give way to all others, while ship 2 must give way to ships 1 and 4.
	As shown in Fig. \ref{fig:cross41_2} - \ref{fig:cross41_3}, the situation is solved according to the traffic rules.

	In the 4 ships scenario, we also compare the performance of the Sync-CCAS and Async-CCAS algorithm with different values of maximum iteration $s_{\max}$.
	As illustrated in Table \ref{table:sync-ccas}, the maximum and mean values of the computation time of an iteration are barely changed when $s_{\max}$ is increased.
	Furthermore, Fig. \ref{fig:smax} shows the similar safety indices between $s_{\max}=2$ and $s_{\max}=5$.
	However, there is a slight decrease of safety indices in case $s_{\max}=10$.
	Taking into account that an increase in the value of $s_{\max}$ would require more computation time (in total) but give no improvement in performance, we can conclude that $s_{\max}=2$ is the optimal value for $s_{\max}$.

	The 6-ships scenario (Fig. \ref{fig:cross61} and \ref{fig:cross62}) is more complex, where multiple ships have the stand-on priority, e.g., ships 1, 4, 5, 6.
	Therefore, negotiation is required to resolve the situation.
	The maximum iteration in this scenario is $s_{\max}=2$.
	Fig. \ref{fig:cross61} presents the result of the Sync-CCAS, while the result of the Async-CCAS is presented in Fig. \ref{fig:cross62}.
	The ships with give-way obligations (ships 2 and 3) strictly follow the traffic rules.
	Ship 2 gave way to all others except ship 3, and ship 3 gave way to all others (Fig. \ref{fig:cross61_4} and \ref{fig:cross62_4}).
	However, there is a difference in the behaviors between ships with stand-on priorities.
	In the synchronous communication scheme, ship 1 gave way to ship 4 (Fig. \ref{fig:cross61_3}).
	While in the asynchronous communication scheme, ship 4 gave way to ship 1 (Fig. \ref{fig:cross62_3}).
	This is because the give-way or stand-on action between ships with the same stand-on priorities is up to negotiation, and the negotiation result is influenced by the communication scheme.
	It is also shown that the order in which ships update their intentions can affect the negotiation outcome between ships with the same stand-on priorities.
	
	Although both Sync-CCAS and Async-CCAS algorithms can resolve the situation without any significant collision risk, the computation time is increased compared to the 4-ships scenarios (see Table \ref{table:sync-ccas}).
	However, these calculation times are acceptable.
	It is worth noting that the average computational times of the Sync-CCAS and Async-CCAS algorithms are similar.

	We further evaluate the performance of the Async-CCAS in the condition of unstable communication between ships.
	The traffic situation setup is the same as in Fig. \ref{fig:cross61}.
	However, each ship $i$ has $5\%$ that its update $\hat{\xi}_i^{s} $ cannot reach other neighboring ships, e.g. ship $i$ lost connection for a short time.
	Due to the data loss, both the maximum and average computation time increase (see Table \ref{table:sync-ccas}).
	Moreover, the safety performance in case of data loss is also decreased (Fig. \ref{fig:loss5}) compared to the case that has no data loss (Fig. \ref{fig:cross62_5}).

	\subsection{Field experiments}
	We verify the performance of the Async-CCAS algorithm with field experiments in the harbor area near Nyhavna, Trondheim, Norway.
	Three Autonomous Surface Vessels (ASVs) called NTNU Fish Otters (see Fig. \ref{fig:otter}) are used in the field experiments.
	%
	%
	Each NTNU Fish Otter is equipped with a computation unit with the core of the Raspberry Pi CM4108032.
	Additionally, the ASVs communicate with each other through a 4G connection.
	Further details of the hardware information of the NTNU Fish Otters can be found in \cite{lauvas_design_2022}.

	In field experiments, the Async-CCAS algorithm is designed to run on the LSTS toolchain \cite{pinto_lsts_2013} and is executed by the onboard computer.
	%
	%
	The update signal $\hat{\xi}_j^{s}$ is exchanged between ships through the IMC communication protocol.
	Each ASV performs the local update every $5$ seconds, with the maximum iteration $s_{\max}=2$.
	%
	%
	To reduce the computation time, we generate a customized solver for the problem \eqref{ADMM-update} using the Casadi codegen tool.
	Then, the parameters, i.e., the initial state $p_{j,i}^\mathrm{init}$, are passed directly to the solver.
	%
	\begin{table}
		\begin{center}
			\begin{threeparttable}
				\caption{Computational time (in second) needed to perform an update in field experiments}
				\label{table:async-ccas}
				{\renewcommand{\arraystretch}{1.25}
					\begin{tabular}{ |c |c|c|c|c| } 
						\hline
						\multicolumn{2}{|c|}{Scenario} & Ship 1 & Ship 2 & Ship 3\\
						\hline
						\multirow{2}{4em}{Fig. \ref{fig:ho}}& Max & 1.7254 & 1.1282 & -\\
						\cline{2-5}
						& Mean & 1.0017 & 0.7109 & -\\
						\cline{2-5}
						\cline{2-5}
						\hline
						\multirow{2}{4em}{Fig. \ref{fig:cross2}}& Max & 1.5899 & 0.4273 & -\\
						\cline{2-5}
						& Mean & 0.8674 & 0.211 & -\\
						\cline{2-5}
						\cline{2-5}
						\hline
						\multirow{2}{4em}{Fig. \ref{fig:ovt}}& Max & 1.0464 & 0.8456 & -\\
						\cline{2-5}
						& Mean & 0.5177 & 0.3996 & -\\
						\cline{2-5}
						\cline{2-5}
						\hline
						\multirow{2}{4em}{Fig. \ref{fig:cross32}}& Max & 1.4445 & 2.1425 & 1.4131\\
						\cline{2-5}
						& Mean & 0.9967 & 1.4406 & 0.457\\
						\cline{2-5}
						\cline{2-5}
						\hline
						\multirow{2}{4em}{Fig. \ref{fig:cross31}}& Max & 0.8734 & 2.5461 & 2.4428\\
						\cline{2-5}
						& Mean & 0.4221 & 1.3135 & 1.5085\\
						\cline{2-5}
						\cline{2-5}
						\hline
				\end{tabular}}
			\end{threeparttable}
		\end{center}
	\end{table}
	\subsubsection{Two-ships scenarios}
\begin{figure}[!t]
			\centering
			\subfloat[\label{fig:ho_1}]{
				\centering
				\includegraphics[width=0.5\linewidth]{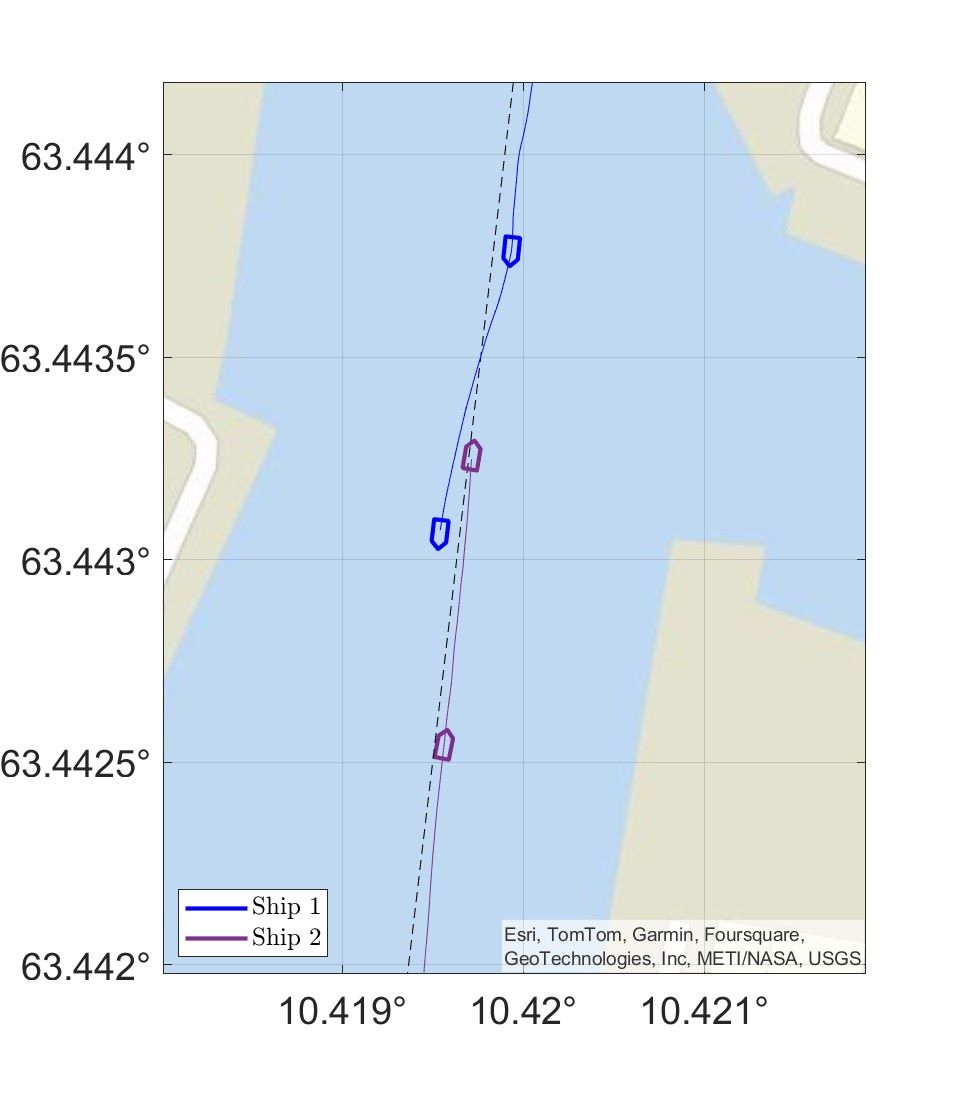}}
			\hfill
			\centering
			\subfloat[Distance between two ships\label{fig:ho_2}]{
				\centering
				\includegraphics[width=0.6\linewidth]{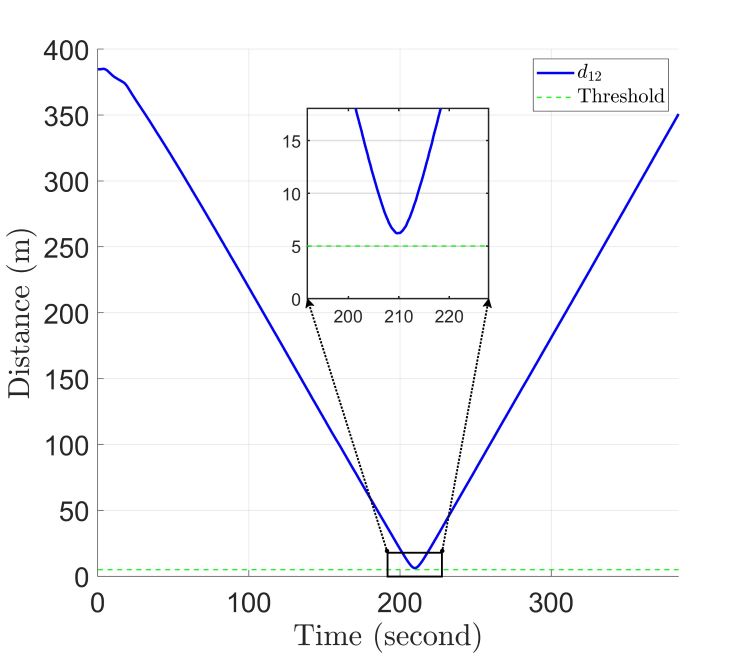}}
			\hfill
			\caption{Field experiments: Head-on between 2 ships. Ship 2 has the stand-on priority. The guiding lines are depicted in the black dashed lines.}
			\label{fig:ho}
		\end{figure}

\begin{figure}[!t]
			\centering
			\subfloat[\label{fig:cross2_1}]{
				\centering
				\includegraphics[width=0.5\linewidth]{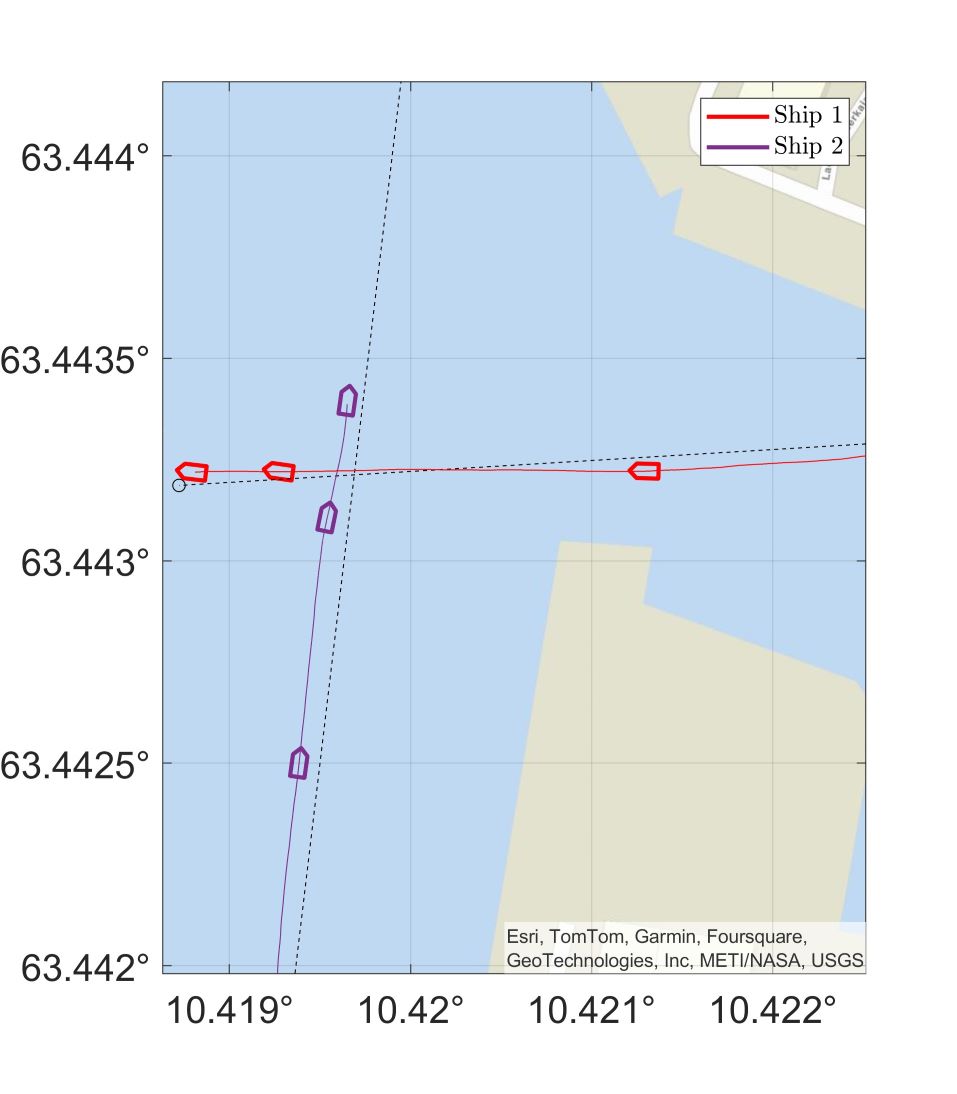}}
			\hfill
			\centering
			\subfloat[Distance between two ships\label{fig:cross2_2}]{
				\centering
				\includegraphics[width=0.6\linewidth]{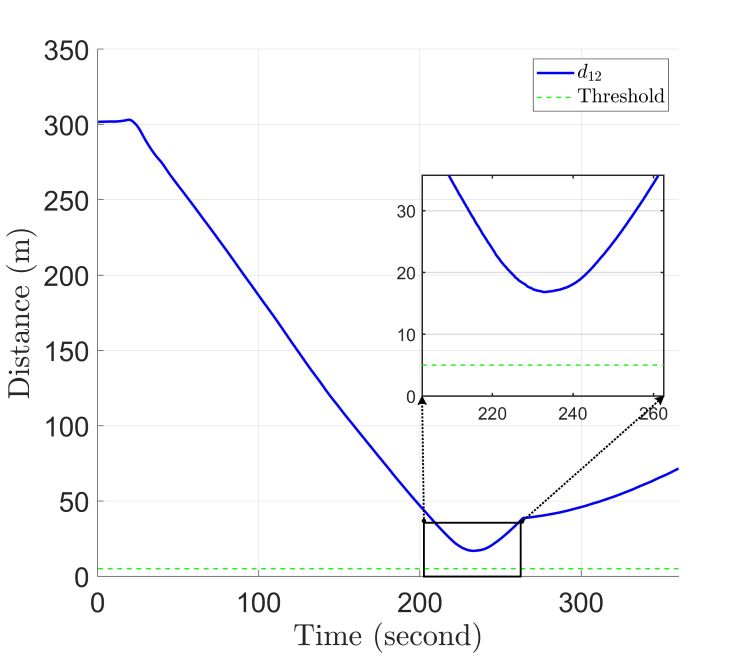}}
			\hfill
			\caption{Field experiments: Intersection crossing between 2 ships. Ship 1 has the stand-on priority. The guiding lines are depicted in the black dashed lines.}
			\label{fig:cross2}
		\end{figure}

\begin{figure}[!t]
			\centering
			\subfloat[\label{fig:ovt_1}]{
				\centering
				\includegraphics[width=0.5\linewidth]{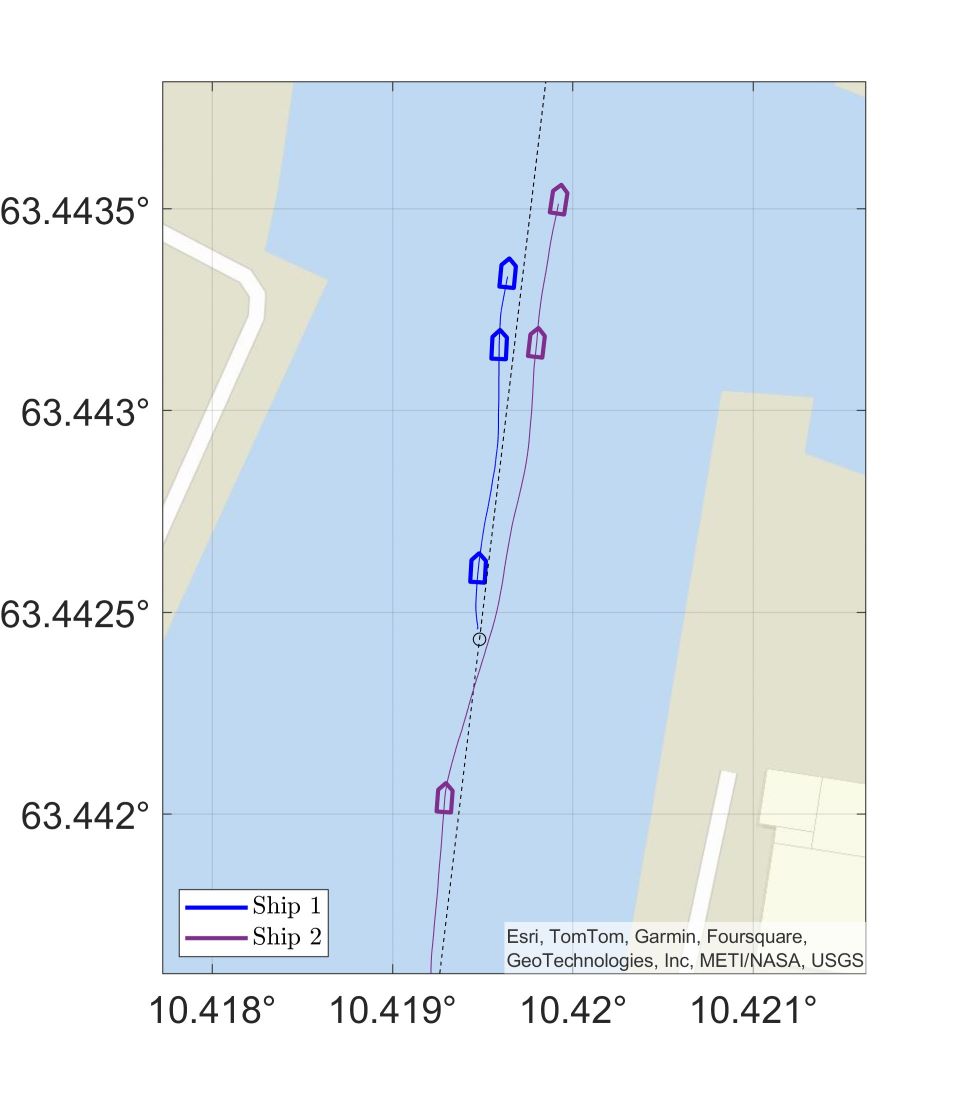}}
			\hfill
			\centering
			\subfloat[Distance between two ships\label{fig:ovt_2}]{
				\centering
				\includegraphics[width=0.6\linewidth]{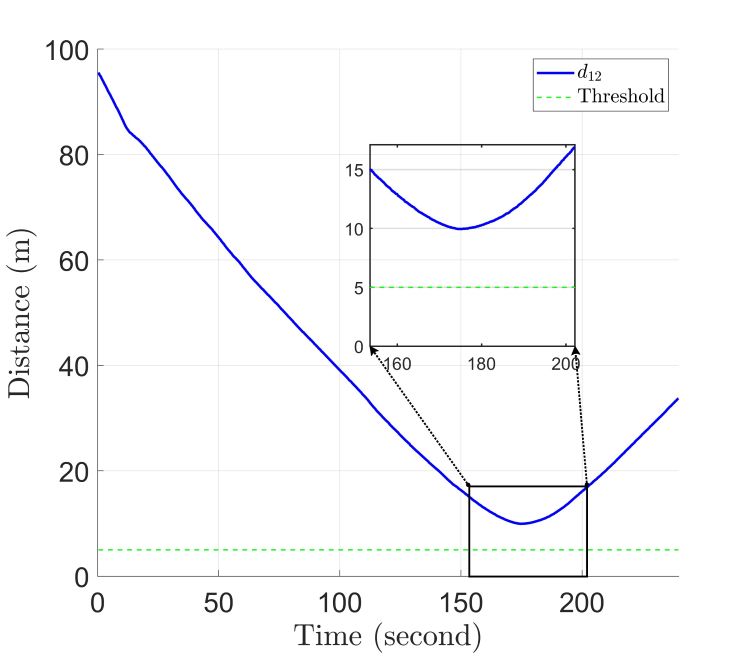}}
			\hfill
			\caption{Field experiments: Overtaking between 2 ships. Ship 1 was overtaken by ship 2. The guiding lines are depicted in the black dashed lines.}
			\label{fig:ovt}
		\end{figure}

	In the first part of the experiment, we verify the performance of Algorithm \ref{async-CCAS} in scenarios with two ships, which are prevalent traffic scenarios.
	The result of head-on, crossing, and overtaking between two ships are shown in Fig. \ref{fig:ho}, \ref{fig:cross2}, and \ref{fig:ovt}, respectively.
	In three scenarios, both stand-on and give-way ships behaved appropriately. 
	In each scenario, the give-way ship changed course (head-on and overtaking scenarios) or reduced speed (crossing scenario) to give way to the stand-on ship.
	\subsubsection{Three-ships scenarios}
\begin{figure*}[!t]
			\centering
			\subfloat[\label{fig:cross32_1}]{
				\centering
				\includegraphics[width=0.25\linewidth]{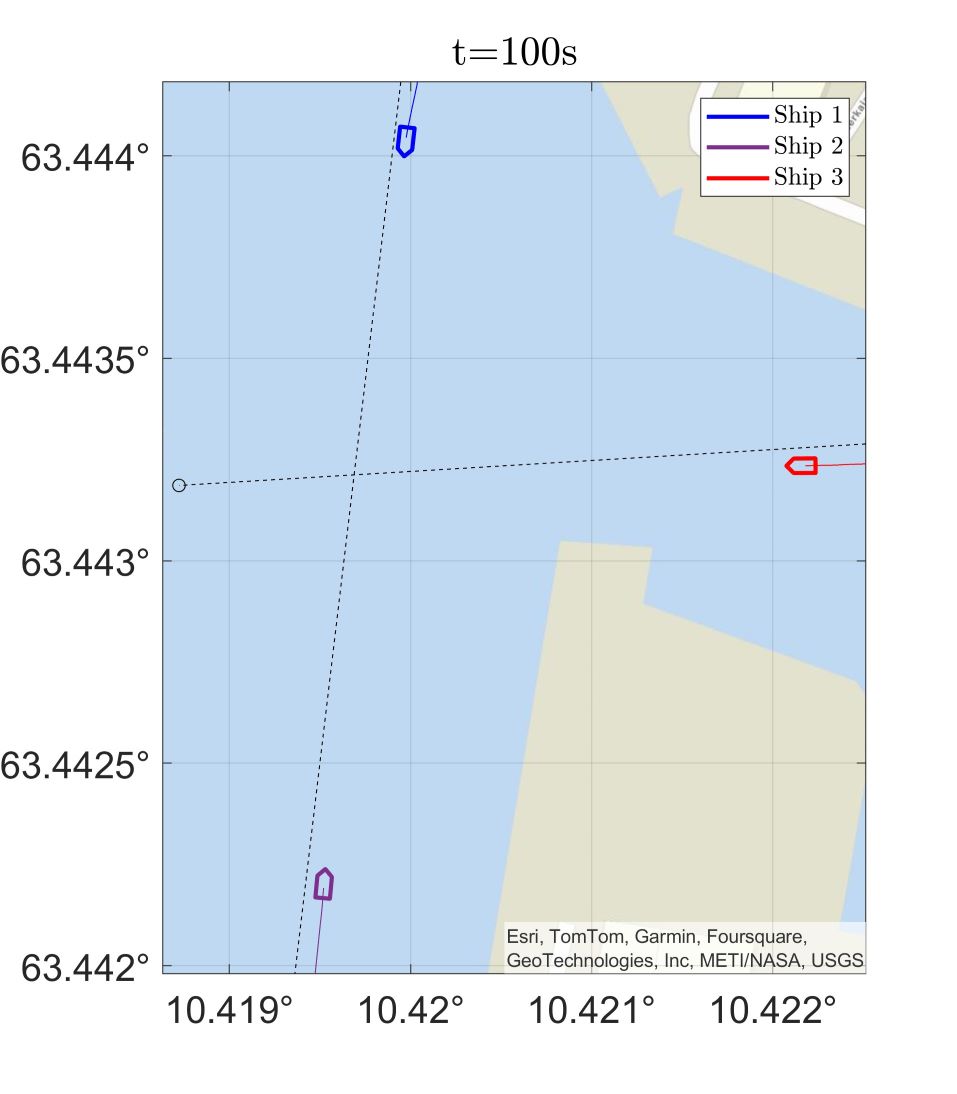}}
			\centering
			\subfloat[\label{fig:cross32_2}]{
				\centering
				\includegraphics[width=0.25\linewidth]{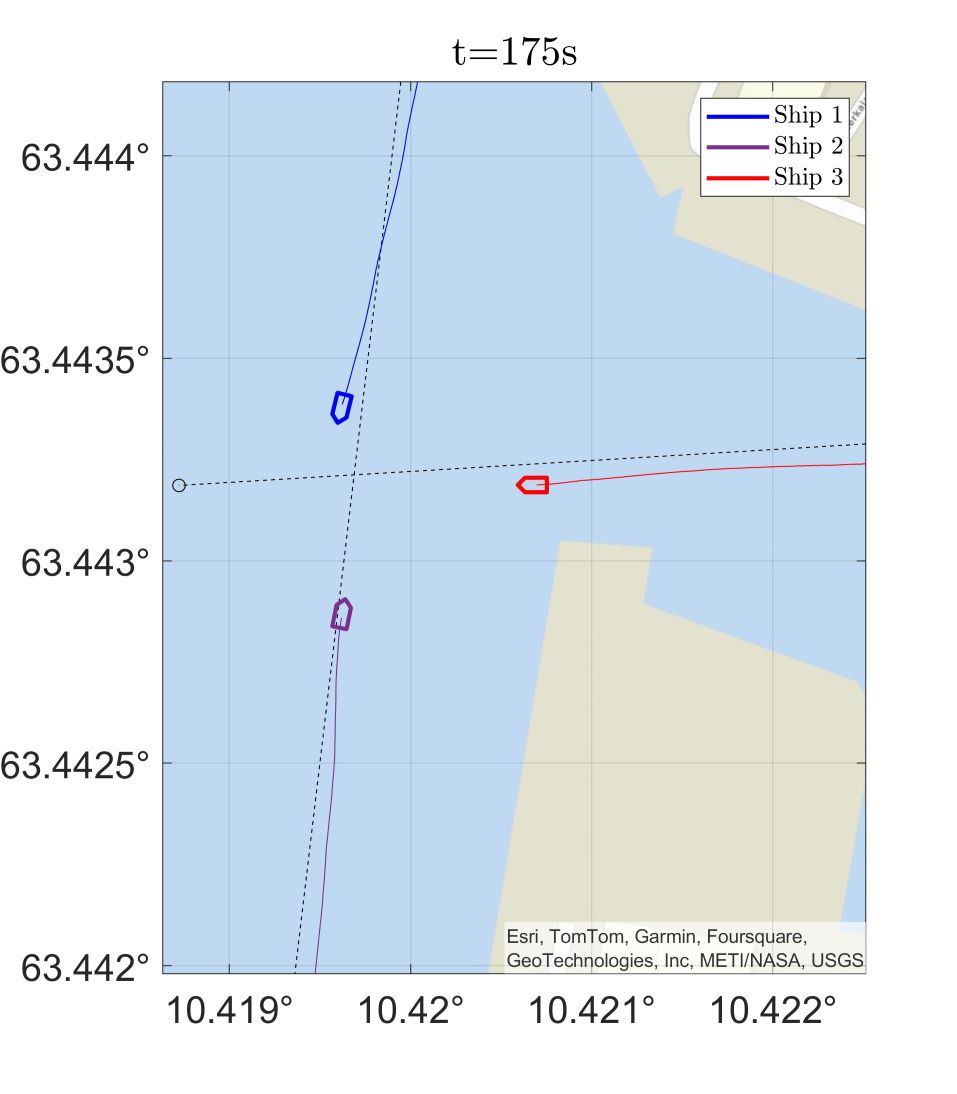}}
			\centering
			\subfloat[\label{fig:cross32_3}]{
				\centering
				\includegraphics[width=0.25\linewidth]{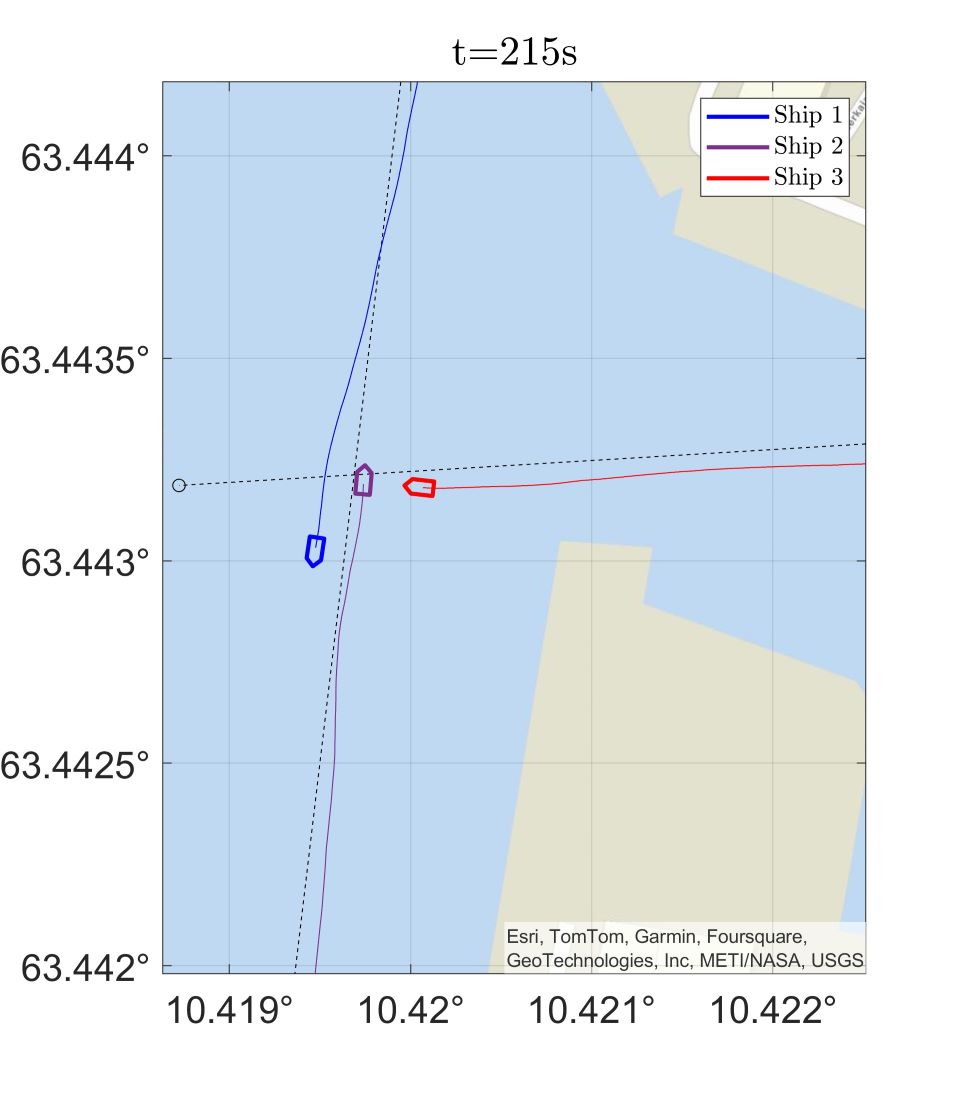}}
			\hfill
			\subfloat[\label{fig:cross32_4}]{
				\centering
				\includegraphics[width=0.25\linewidth]{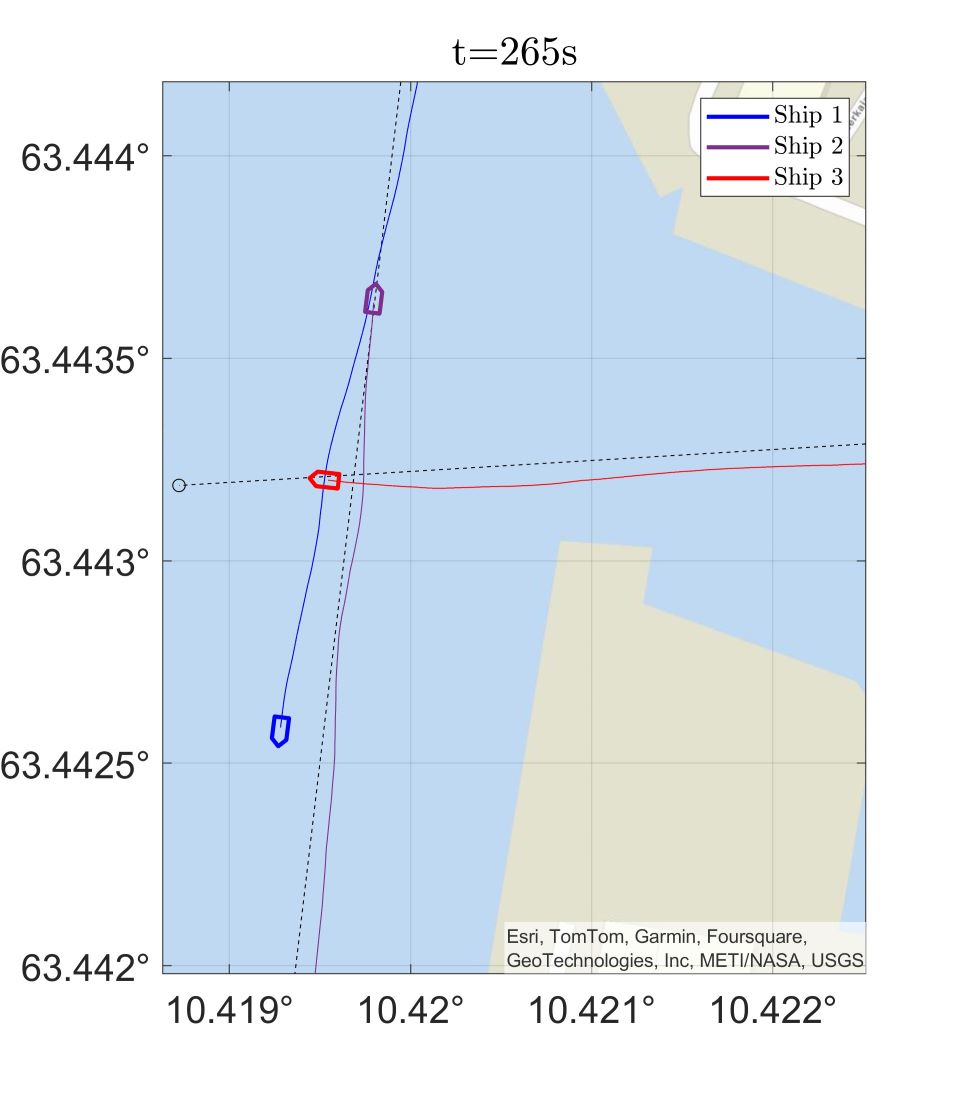}}
			\centering
			\subfloat[Distance between ships\label{fig:cross32_5}]{
				\centering
				\includegraphics[width=0.3\linewidth]{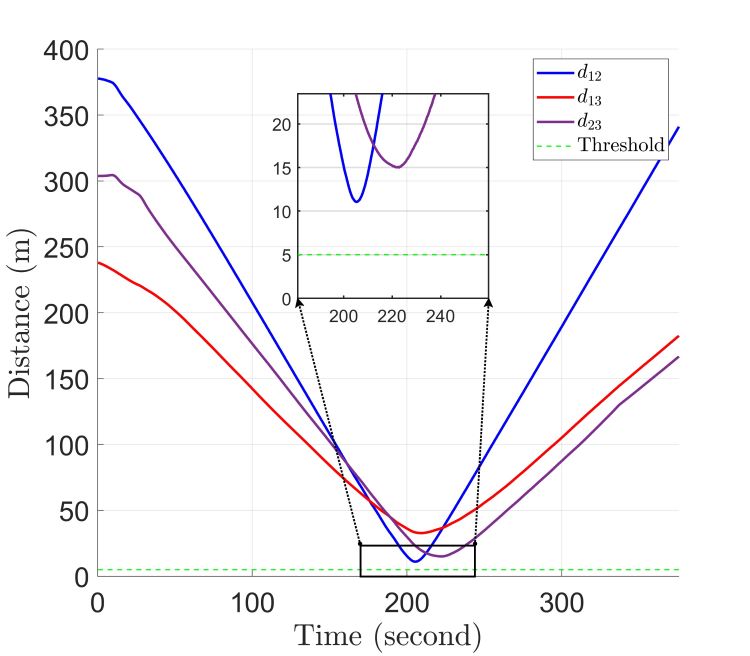}}
			\caption{Field experiments: Intersection crossing between 3 ships. The guiding lines are depicted in the black dashed lines.}
			\label{fig:cross32}
		\end{figure*}

\begin{figure*}[!t]
			\centering
			\subfloat[\label{fig:cross31_1}]{
				\centering
				\includegraphics[width=0.25\linewidth]{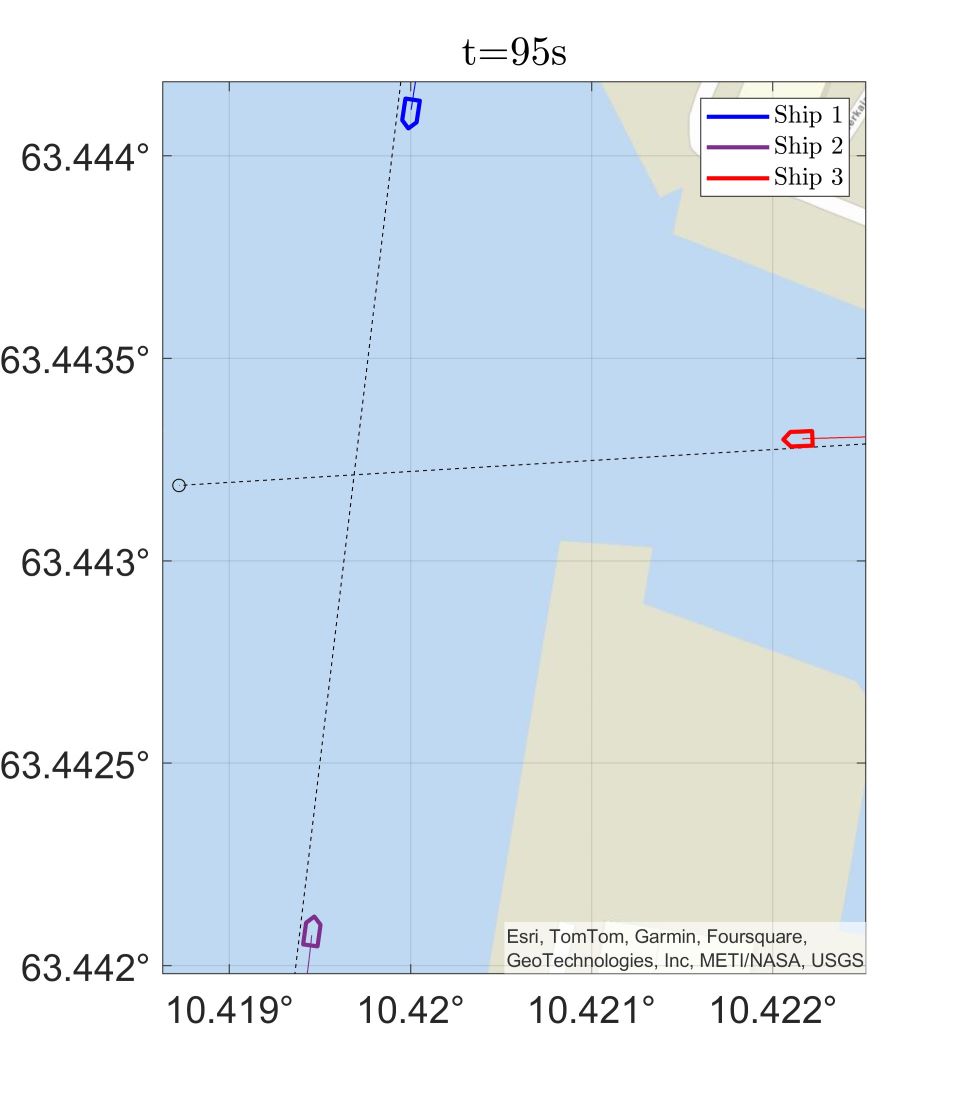}}
			\centering
			\subfloat[\label{fig:cross31_2}]{
				\centering
				\includegraphics[width=0.25\linewidth]{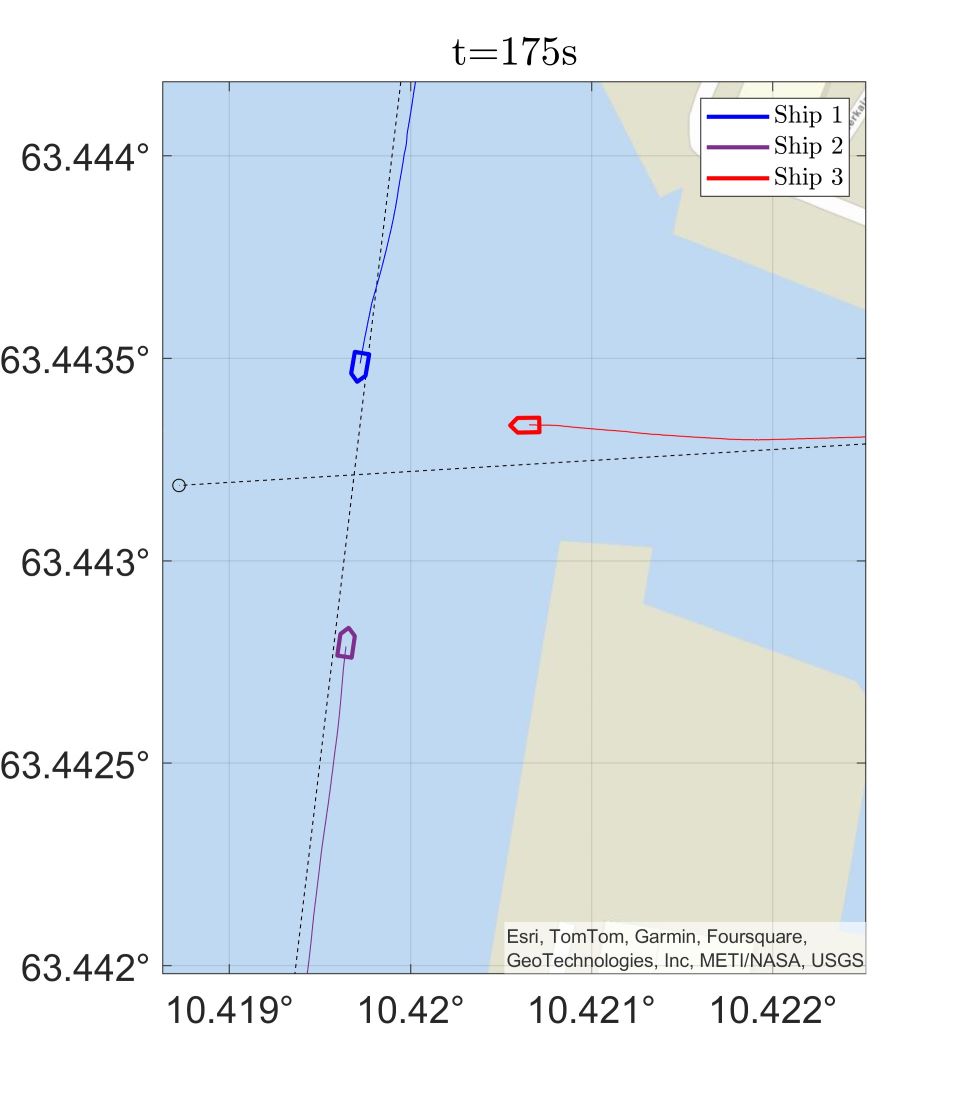}}
			\centering
			\subfloat[\label{fig:cross31_3}]{
				\centering
				\includegraphics[width=0.25\linewidth]{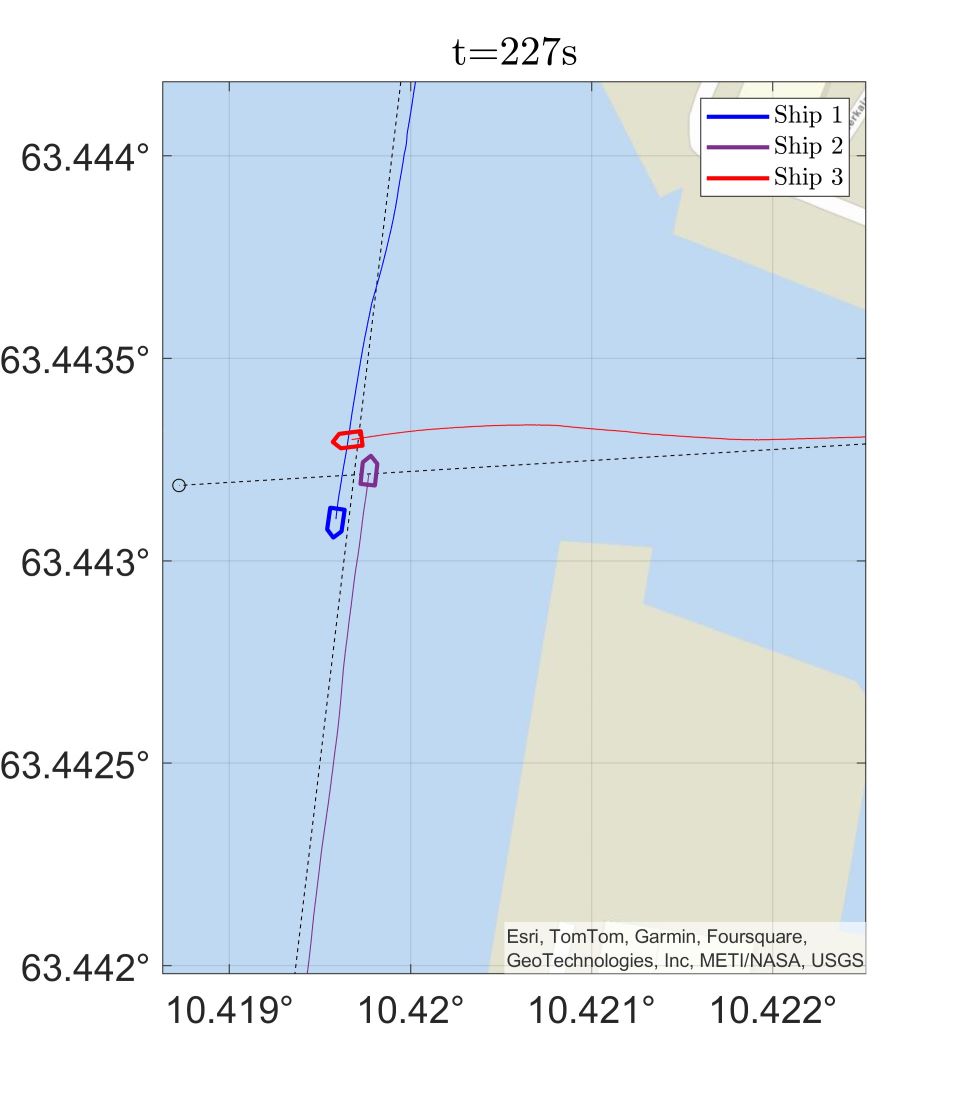}}
			\hfill
			\subfloat[\label{fig:cross31_4}]{
				\centering
				\includegraphics[width=0.25\linewidth]{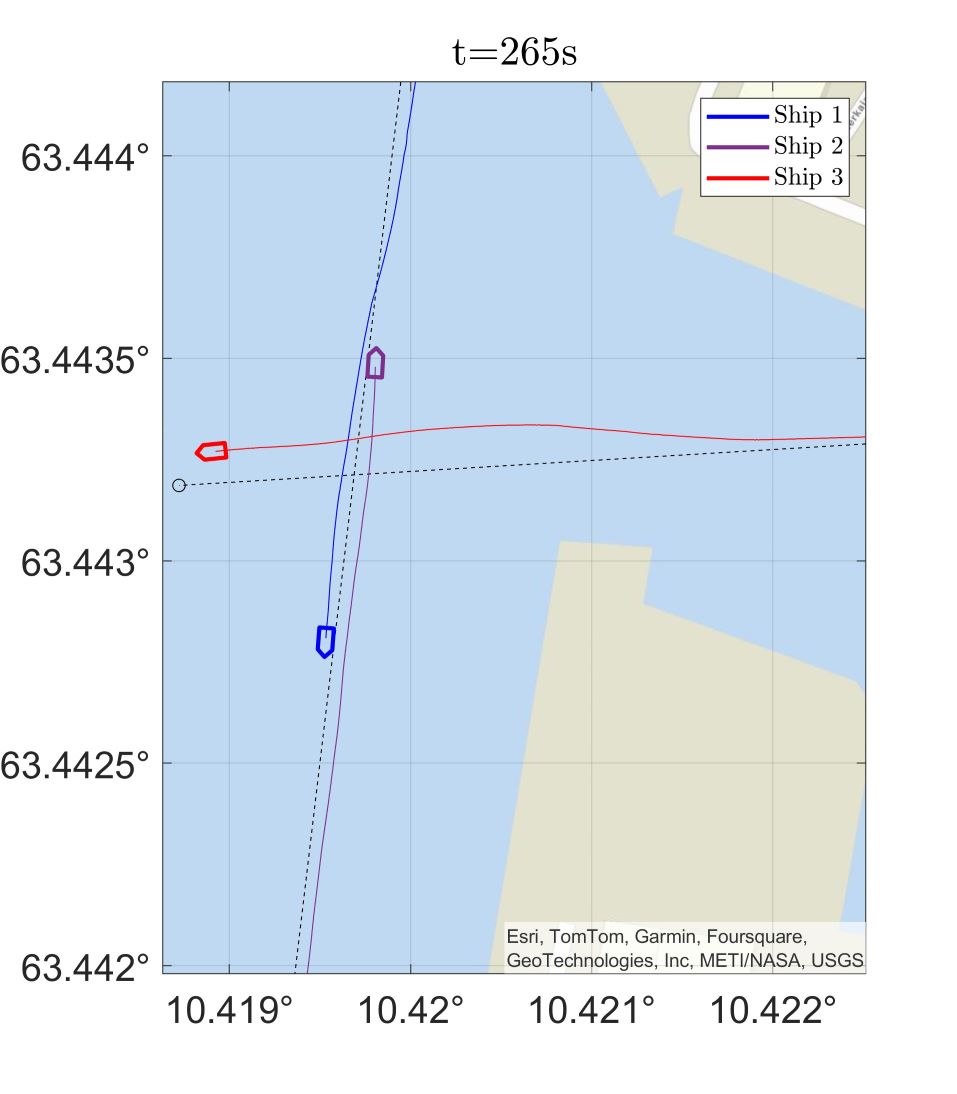}}
			\centering
			\subfloat[Distance between ships\label{fig:cross31_5}]{
				\centering
				\includegraphics[width=0.3\linewidth]{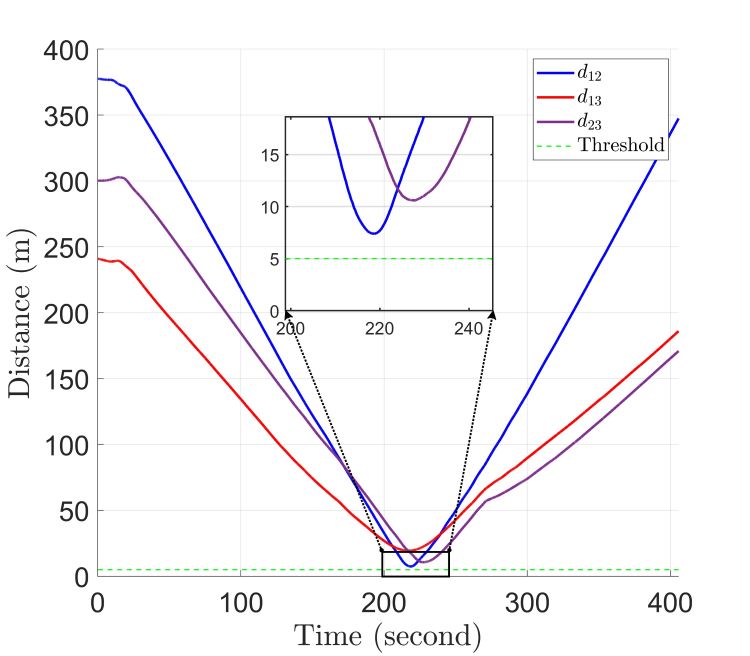}}
			\caption{Field experiments: Intersection crossing between 3 ships. The guiding lines are depicted in the black dashed lines.}
			\label{fig:cross31}
		\end{figure*}

	The second part of the experiment involves three ships in combined head-on and crossing scenarios (see Fig. \ref{fig:cross32} and \ref{fig:cross31}).
	In both scenarios, ships 1 and 2 encounter each other head-on, and simultaneously, ship 3 passes by from the east.
	The only difference between the two scenarios is that in the first one (Fig. \ref{fig:cross32}), there is a clear stand-on and give-way order between the three ships.

	In the first scenario (Fig. \ref{fig:cross32_1}), ship 3 must give way for ships 1 and 2 because it is sailing on the port sail of the guiding line.
	Ship 1 must give way to ship 2, and ship 2 can stand on.
	With a clear priority order, Algorithm \ref{async-CCAS} resolved the situation as shown in Fig. \ref{fig:cross32_3} and \ref{fig:cross32_4}.

	In the second scenario (Fig. \ref{fig:cross31_1}), ships 2 and 3 sail on the starboard side and, therefore, can stand on.
	In this scenario, only ship 1 must give way to ships 2 and 3.
	This is a situation where the priority by traffic rules is unclear since ships 2 and 3 both have stand-on priority.
	Therefore, a negotiation must be carried out between ships 2 and 3.
	As illustrated in  Fig. \ref{fig:cross31_5}, the minimum distance between ships 2 and 3 is approximately $10\mathrm{m}$ and significantly lower than that of the first scenario, which is $15\mathrm{m}$ (see Fig. \ref{fig:cross32_5}).
	Additionally, from Table \ref{table:async-ccas}, the mean and maximum calculating time of ship 3 in Fig. \ref{fig:cross32} is higher than in Fig. \ref{fig:cross31}.
	The increase in calculating time and decrease in distance is due to the unclear priority between ships 2 and 3, which require more negotiation time.
	Therefore, evasive actions were executed later.
	Nevertheless, the minimum distances in both scenarios are above the safety threshold.
		%
		%
		%
		
		\section{Conclusion and future research}\label{sec:conclusion}
		This paper presented two distributed MPC-based ADMM algorithms to solve the problem of collaborative collision avoidance for autonomous ships in inland waterways.
		The proposed algorithms allow ships to collaborate to avoid collisions and comply with inland traffic regulations.
		On the one hand, the synchronous algorithm is designed for an ideal condition in which the information exchange process is synchronized between ships.
		On the other hand, the asynchronous algorithm is designed to handle cases where synchronized communication is not available.
		We overcome the obstacle of asynchronous communication by extending the result of the asyncFedDR.

		The simulation results show that the synchronous algorithm can guarantee collision-free navigation for ships in complex scenarios.
		In the simulation environment, the synchronous and asynchronous algorithms show similar results and computational times.
		Both algorithms comply with traffic rules in tested scenarios.
		Furthermore, field experiments confirm the performance of the synchronous algorithm under practical conditions.
		However, the simulation results suggest that the proposed algorithms' performance can decrease in more extreme conditions, such as unstable communication between ships.
		
		Although the proposed algorithm can handle cases where the priority based on traffic rules is unclear, computation time is increased. 
		Future research could aim to reduce computation time in these cases. 
		\section*{Acknowledgment}
		We would like to thank Melih Akda\u{g} for providing the code template that appears as a foundation for the collision avoidance algorithm in a field experiment.
		We would also like to express our gratitude to Daniel Bogen, Jo Arve Alfredsen, Lorenzo Govani, and Morten Einarsve for your invaluable support during the field experiment.
		\appendices
		\section{Proof of Lemma \ref{ADMM_requirement}}
		\label{lemma-proof}
		$\bullet$ Lemma \ref{ADMM_requirement}\ref{lemma:lsc}.
		$f_i(w_i)$ is a sum of an indicator function of a closed set $\Gc_i(\tilde{p}_{i},\tilde{u}_i)$ and a continuous function $\Jc_i(\tilde{p}_{i},\tilde{u}_{i})$.
		Thus, $f_i(w_i)$ is lsc.
		Since $g(\xi)=0$, then $g$ is proper, closed and convex.
		
		$\bullet$ Lemma \ref{ADMM_requirement}\ref{lemma:surjective}. $A_i$ is a full-row rank matrix and is then surjective. Hence, $A$ is surjective.
		
		$\bullet$ Lemma \ref{ADMM_requirement}\ref{lemma:L-smooth}.
		First, we show that there exist $L_{A_if_i}>0$ such that
		\begin{align}\label{lemma:L-smooth-con}
			-L_{A_if_i} ||A_i(w_i^1-w_i^2)||^2 &\leq \langle \nabla f_i(w_i^1) -\nabla f_i(w_i^2),w_i^1-w_i^2 \rangle \nonumber \\
			&\leq L_{A_if_i} ||A_i(w_i^1-w_i^2)||^2,
		\end{align}
		whenever $\nabla f_i(w_i^1),\nabla f_i(w_i^1) \in \textbf{range}(A_i^\top)$.
		%
		%
		Then $\nabla f_i(w_i)\in \textbf{range}(A_i^\top)$ is equivalent to $\tilde{u}_i = 0$ and $\Gc_i =0$. 
		In that case, $f_i(w_i) = \Jc_i(\tilde{p}_{i} ,0)= \Jc_i^{ca}(\tilde{p}_i)$ is Lipschitz continuous gradient function due to $R_{ij}(\cdot)$ being $L_R$-smooth.
		Then there exist $L_{f_i}$ such that
		\begin{align*}
			||\nabla f_i(w_i^1) -\nabla f_i(w_i^2)|| \leq  L_{f_i} ||w_i^1-w_i^2||
		\end{align*}
		With the help of the Cauchy-Schwartz inequality, we have
		\begin{align*}
			L_{f_i} ||w_i^1-w_i^2||^2 &\geq ||\nabla f_i(w_i^1) -\nabla f_i(w_i^2)||.||w_i^1-w_i^2|| \\
			&\geq |\langle \nabla f_i(w_i^1) -\nabla f_i(w_i^2),w_i^1-w_i^2 \rangle|,
		\end{align*}
		or equivalently 
		\begin{align}\label{lemma:1condition}
			-L_{f_i} ||w_i^1-w_i^2||^2 &\leq \langle \nabla f_i(w_i^1) -\nabla f_i(w_i^2),w_i^1-w_i^2 \rangle \non\\
			&\leq L_{f_i} ||w_i^1-w_i^2||^2,
		\end{align}
		We recall fact that $\tilde{R}_i\tilde{p}_i - c_i = \xi$, and $R_i$, which is the component of $\tilde{R}_i$, is the rotation matrix.
		Therefore $||\tilde{p}_i^1-\tilde{p}_i^2||^2 = ||\xi^1-\xi^2)||^2$.
		Then for all $\nabla f_i(w_i)\in \textbf{range}(A_i^\top)$ we have
		\begin{align}\label{lemma:2condition}
			L_{f_i} ||w_i^1-w_i^2||^2 &= L_{f_i} ||\tilde{p}_i^1-\tilde{p}_i^2||^2 \non\\
			&= L_{f_i} ||\xi^1-\xi^2||^2 \non\\
			&=  L_{f_i} ||A_i(w_i^1-w_i^2)||^2
		\end{align}
		From \eqref{lemma:1condition} and \eqref{lemma:2condition} we get there exist $L_{A_if_i}$ satisfies \eqref{lemma:L-smooth-con}.
		Then, following \cite[Thm 5.13]{themelis_20}, $(A_if_i)$ is $L_(A_if_i)$-smooth.
		Furthermore, if we choose $L$ such that $L = \max_{i\in\Mc}\{L_{(A_i f_i)}\}$, we also have
		\begin{align*}
			-L ||A(w^1-w^2)||^2 &\leq \langle \nabla f(w^1) -\nabla f(w^2),w^1-w^2 \rangle \nonumber \\
			&\leq L ||A(w^1-w^2)||^2,
		\end{align*}
		holds whenever $\nabla f(w^1),\nabla f(w^1) \in \textbf{range}(A^\top)$.
		Therefore, $(Af)$ is $L$-smooth.
		\qed
		\section{Proof of Theorem \ref{ADMM equivalent}}
		\label{R-ADMM-proof}
		First, we show that \eqref{admm-drs-update} holds.
		Follows \eqref{drs-admm:var}, we have
		\begin{align}
			\varsigma_i^{s} \!+ \!\lambda(\tau^{s}-u_i^{s}) &= A_i w_i^{s} -c_i - \frac{z_i^{s}}{\beta} - \lambda(A_i w_i^{s} -v^{s} -c_i)\non \\
			&\myeq{\eqref{z0.5}} A_i w_i^{s} \!-\! c_i - \frac{z_i^{s+0.5}}{\beta} - (A_i w_i^{s} \!-\! v^{s} \!-\! c_i) \non \\
			&\myeq{\eqref{z1}} A_i w_i^{s+1} -c_i- \frac{z_i^{s+1}}{\beta} = \varsigma_i^{s+1}. \label{s+}
		\end{align}
		Moreover, from \eqref{s+} we also observer that
		\begin{align}
			\varsigma^{s+1}_i = -\frac{z_i^{s+0.5}}{\beta} - v^s. \label{s-z}
		\end{align}
		Next, we have
		\begin{align}
			\textbf{prox}_{\gamma\varphi_i} (\varsigma_i^{s+1}) &= \arg \min_{\mu}\Bigl\{(A_if_i)(\mu+c_i) \nonumber\\
			&~~~~~~~~~~~~~~~~~~+\frac{1}{2\gamma}||\mu-\varsigma_i^{s+1}||^2\Bigr\} \nonumber\\
			&= \arg \min_{\mu}\Bigl\{\inf_{w_i} f_i(w_i)  \nonumber \\
			&~~~~~~~~~+\frac{1}{2\gamma}||\mu-\varsigma_i^{s+1}||^2\big|A_iw_i=\mu+c_i\Bigr\} \nonumber\\
			&=A_i\arg \min_{w_i}\Bigl\{f_i(w_i) \nonumber \\
			&~~~~~+\frac{1}{2\gamma}||A_iw_i-c_i-\varsigma_i^{s+1}||^2\Bigr\} -c_i. \label{u-z}
		\end{align}
		Then, we obtain:
		\begin{align*}
			u_i^{s+1} &= A_iw_i^{s+1} -c_i \\
			&=A_i\arg \min_{w_i}\Bigl\{f_i(w_i)+\left\langle z_i^{s+0.5}, A_i w_i -v^{s} -c_i\right\rangle \\
			&~~+\frac{\beta}{2}||A_i w_i +v^{s} -c_i||^2\Bigr\} -c_i \\
			&= A_i\arg \min_{w_i}\Bigl\{f_i(w_i) \\
			&~~+\frac{\beta}{2}||A_i w_i +v^{s} -c_i+\frac{z_i^{s+0.5}}{\beta}||^2\Bigr\} -c_i \\
			&\myeq{\eqref{s-z}} A_i\arg \min_{w_i}\Bigl\{f_i(w_i) \\ 
			&~~~~~~~~~~~~+\frac{\beta}{2}||A_i w_i-c_i-\varsigma_i^{s+1}||^2\Bigr\}-c_i \\
			&\myeq{\eqref{u-z}} \textbf{prox}_{\gamma\varphi_i} (\varsigma_i^{s+1}).
		\end{align*}
		Similarly
		\begin{align*}
			\tau^{s+1} &= v^{s+1} \\
			&= \arg \min_{v}\biggl\{g(v)+\Bigl\langle \frac{1}{M}\sum_{i=1}^{M} z_i^{s+1}, \tilde{w}^{s+1}-v \Bigr \rangle\\
			&~~~~~~~~~~~~~~~~ +\frac{\beta}{2}||\tilde{w}^{s+1}-v||^2\biggr\} 
		\end{align*}
		\begin{align*}
			&= \arg \min_{v}\biggl\{g(v) \\
			&~~~~~~~ +\frac{\beta}{2}||\frac{1}{M}\sum_{i=1}^{M}\bigl(A_i w_i^{s+1} -c_i+\frac{z_i^{s+1}}{\beta} \bigr)-v||^2\biggr\}  \\
			&= \arg \min_{v}\biggl\{g(v) +\frac{\beta}{2}||v-\frac{1}{M}\sum_{i=1}^{M}(2u_i^{s+1}-\varsigma_i^{s+1})||^2\biggr\}  \\
			&= \textbf{prox}_{\gamma \phi}\left\{\frac{1}{M}\sum_{i=1}^{M}(2u_i^{s+1}-\varsigma_i^{s+1})\right\}.
		\end{align*}
		Now, to prove \eqref{dre}, we first show that:
		\begin{align}
			\nabla \varphi_i(u_i^{s}) = -z_i^s,
		\end{align}
		hold for all $i \in \Mc$. Indeed, since
		\begin{align*}
			\begin{split}
				u_i^{s} &= \textbf{prox}_{\gamma\varphi_i} (\varsigma_i^{s}) ~~\forall i \in \Mc, \\
				z_i^{s} &= -\frac{\varsigma_i^s-u_i^s}{\gamma},
			\end{split}
		\end{align*}
		and together with \eqref{prox-con} imply that
		\begin{align}\label{dre1}
			z_i^{s} &= -\frac{\varsigma_i^s-u_i^s}{\gamma} = -\nabla \varphi_i(u_i^s).
		\end{align}
		Next, \cite[Prop 5.2(ii)]{themelis_20} gives:
		\begin{align}\label{dre2}
			\varphi(u_i^s)&=(A_i f_i)(A_i w_i^s) = f_i(w_i^s).
		\end{align}
		From \eqref{dre1} and \eqref{dre2} then \eqref{dre} holds.\qed
		
		\bibliographystyle{IEEEtran}
		\bibliography{mybibfile,MyEndnoteLibrary}
		
		
	\end{document}